%% file: dynamate-journal.tex
\newif\ifarXiv
\arXivtrue

\documentclass[10pt,journal,cspaper,compsoc]{IEEEtran}

\usepackage[pdftex]{graphicx}

\usepackage[cmex10]{amsmath}
\usepackage{algorithmic}

\usepackage[tight,normalsize,sf,SF]{subfigure}

\usepackage{url}

\usepackage[normalem]{ulem}
\usepackage{amstext}
\usepackage{xspace}
\usepackage{alltt}
\usepackage{color}
\usepackage{boxedminipage}
\usepackage{hyperref}

\usepackage{mathtools}
\usepackage{booktabs}
\usepackage{algorithm}
\usepackage{paralist}
\usepackage{multirow}
\usepackage[colorinlistoftodos,textsize=tiny]{todonotes}

\usepackage{flushend}
\usepackage{latexsym}

\usepackage{changepage}
\usepackage{tikz}
\usetikzlibrary{shapes,arrows,positioning,calc}

\usepackage{listings}
\input{jml-listings}
\usepackage[T1]{fontenc}
\usepackage[scaled=0.81]{beramono} 
\newenvironment{result}%
{\bigskip
\noindent
\let\emph=\textbf
\begin{boxedminipage}{\columnwidth}\begin{center}\em}%
{\end{center}\end{boxedminipage}%
\bigskip
}

\newcommand{\DAIKON}{\textsc{Dai\-kon}\xspace}
\newcommand{\EVOSUITE}{\textsc{EvoSuite}\xspace}
\newcommand{\JAVA}{Ja\-va\xspace}
\newcommand{\JML}{{\small JML}\xspace}
\newcommand{\HOUDINI}{{\small HOUDINI}\xspace}
\newcommand{\HAVOC}{{\small HAVOC}\xspace}
\newcommand{\GUESSANDCHECK}{{\small GUESS-AND-CHECK}\xspace}
\newcommand{\DYSY}{{\small DYSY}\xspace}

\newcommand{\DASH}{{\small DASH}\xspace}
\newcommand{\SYNERGY}{{\small SYNERGY}\xspace}
\newcommand{\JSTAR}{\textsf{jStar}\xspace}

\newcommand{\DYNAMATE}{{\small DYNAMATE}\xspace}
\newcommand{\ESCJ}{{ESC/Ja\-va2}\xspace}
\newcommand{\GINDYN}{\textsc{Gin-Dyn}\xspace}

\newcommand{\HOLA}{\textsc{Hola}\xspace}
\newcommand{\CCC}{\texttt{cc\-check}\xspace}
\newcommand{\BLAST}{\textsc{Blast}\xspace}
\newcommand{\INVGEN}{\textsc{InvGen}\xspace}

\newcommand{\limpl}{\Longrightarrow}

\newcommand{\expr}[1]{\ensuremath{\mathcal{E}_{\text{#1}}}}
\newcommand{\muts}{\ensuremath{\mathcal{M}}}

\newcommand{\nicepar}[1]{\vskip 4pt\textbf{#1}}

\newcommand{\mathid}[1]{\text{\rmfamily\textsl{#1}}}
\def\|#1|{\mathid{#1}}

\def\<#1>{{\lstinline[basicstyle=\normalsize\ttfamily]|#1|}}
\def\!#1!{{\lstinline[basicstyle=\scriptsize\ttfamily]|#1|}}
\newcommand{\mcode}[1]{\text{\<#1>}}

\newif\iflong
\longfalse

\begin{document}

\title{Inferring Loop Invariants by Mutation, Dynamic Analysis, and Static Checking}

\ifarXiv
\author{%
Juan~P.~Galeotti$^1$, 
Carlo~A.~Furia$^2$, 
Eva~May$^1$, 
Gordon~Fraser$^3$, 
and~Andreas~Zeller$^1$

\vspace{2mm}
$^1$ Chair of Software Engineering, Saarland University, Saarbr\"ucken, Germany

$^2$ Chair of Software Engineering, Department of Computer Science, ETH Zurich, Switzerland

$^3$ University of Sheffield, UK
}
\fi

\IEEEcompsoctitleabstractindextext{%
\begin{abstract}
Verifiers that can prove programs correct against their full functional specification require, for programs with loops, additional annotations in the form of \emph{loop invariants}---prop\-er\-ties that hold for every iteration of a loop.  
We show that significant loop invariant candidates can be generated by systematically mutating postconditions; then, dynamic checking (based on automatically generated tests) weeds out invalid candidates, and static checking selects provably valid ones.
We present a framework that automatically applies these techniques to support a program prover, paving the way for fully automatic verification without manually written loop invariants: 
Applied to 28~methods (including 39~different loops) from various \<java.util> classes (occasionally modified to avoid using Java features not fully supported by the static checker), our \DYNAMATE prototype automatically discharged 97\%
of all proof obligations, resulting in automatic complete correctness proofs of 25~out of the 28~meth\-ods---outperforming several state-of-the-art tools for fully automatic verification.
\end{abstract}

}

\maketitle

\section{Introduction}
\label{sec:introduction}

\IEEEPARstart{D}{espite} significant progress in automating program verification, proving a  program correct still requires substantial expert manual effort.  For programs with loops, 
one of the biggest burdens is providing \emph{loop invariants}---properties that hold at the entry of a loop and are preserved by an arbitrary number of loop iterations.
Compared to other specification elements such as pre- and postconditions, loop invariants tend to be difficult to understand and to express, and even structurally simple loops may require non-trivial invariants to be proved correct~\cite{LoopInvariantSurvey-TR-19112012}.

In this paper, we present and evaluate a novel approach to improve the automation of full program verification through the automatic discovery of suitable loop invariants.
The approach is based on a combination of static (program proving) and dynamic (testing) techniques that complement each other's capabilities.
The key observation is that, given loop invariant \emph{candidates} (assertions that may or may not hold for the loops under analysis), an automatic testing tool can efficiently weed out \emph{invalid} candidates, whereas a program verifier can promptly \emph{validate} candidates---and possibly use them to prove the program correct.
Thus, if we can \emph{produce a suitable set of plausible loop invariant candidates}, we have chances to prove the program correct in a fully automatic fashion, without need for additional annotations.

\begin{figure}
\begin{lstlisting}
 private static int binarySearch0(int[] a, 
       int fromIndex, int toIndex, 
       int key) {
   int low = fromIndex, high = toIndex - 1;
   while (low <= high) {
     // midpoint of #[low..high]#
     int mid = low + ((high - low)/2);  
     int midVal = a[mid];
     if (midVal < key) low = mid+1;
      else if (midVal > key) high = mid - 1;
      else return mid; // key found
   }
   return -(low + 1); // key not found
 }
\end{lstlisting}
\caption{Binary search method in \<java.util.Arrays>.}
\label{fig:binary-search-code}
\end{figure}

\begin{figure}
\begin{lstlisting}
/*@
  @ requires a != null;
  @ requires TArrays.within(a, fromIndex, toIndex);
  @ requires TArrays.sorted(a, fromIndex, toIndex);
  @ 
  @ ensures \result >= 0 ==> a[\result] == key;
  @ ensures \result < 0 
  @     ==> !TArrays.has(a, fromIndex, toIndex, key);
  @*/
\end{lstlisting}
\caption{Pre- and postcondition of \<binarySearch0>.}
\label{fig:binary-search-spec}
\end{figure}

\subsection{Running Example: Binary Search}
\autoref{fig:binary-search-code} shows \<binarySearch0>, a helper method declared in class \<java.util.Arrays> in the standard \JAVA API.
Method \<binarySearch0> takes a sorted array \<a>; if element \<key> is found in \<a>, it returns its index, otherwise it returns a negative integer.
\autoref{fig:binary-search-spec} formalizes this behavior as pre- and postcondition written in \JML \cite{leavens2006}, using model-based \emph{predicates}~\cite{PFM10-VSTTE10}, representing implicit quantified expressions, with descriptive names.
For example, the predicate \<!TArrays.has(a,> \<fromIndex,toIndex, key)> means that array \<a> has no element \<key> over the interval range from \<fromIndex> (included) to \<toIndex> (excluded). 
The specification of \autoref{fig:binary-search-spec} is what we want to verify \<binarySearch0> against, and it would be required for any kind of functional validation---be it based on testing or static reasoning. 

Fully automatic verifiers such as \CCC~\cite{DBLP:conf/foveoos/FahndrichL10} (formerly known as Clousot~\cite{CousotCL11}) or \BLAST~\cite{blast} fail to establish the correctness of the annotated program. 
On the other hand, auto-active verifiers such as \ESCJ~\cite{ESCJava2}\footnote{\label{footnote:note-soundness} Like most static verifiers working on real programming languages, \ESCJ does not fully support a number of Java and JML language features, which makes it an unsound tool in general. In this work, we only deal with a subset of Java and JML that excludes \ESCJ's unsupported features, and we always use \ESCJ with the \texttt{-loopSafe} option for sound verification of unbounded loops. Therefore, the rest of the paper refers to \ESCJ as a \emph{sound but incomplete tool}, with the implicit proviso that we avoid exercising its known sources of unsoundness. See \autoref{sec:case-study-preparation} for more details about this issue with reference to the case study.} succeed, but only if we provide suitable invariants for the loop in \autoref{fig:binary-search-code}, such as those in \autoref{fig:binary-search-inv}. 
Compared to pre- and postconditions, it is much more difficult to write loop invariants, since they capture implementation-specific details rather than general input/output behavior.
A method implementing a different search algorithm would have the same postcondition as \<binarySearch0>'s, but proving it correct would most likely require quite different loop invariants.

\begin{figure}[ht]
\centering\scriptsize
\begin{lstlisting}[numbers=left,mathescape]
/*@
  @ loop_invariant fromIndex <= low $\label{l:binarySearch-bounding-inv-1}$
  @ loop_invariant low <= high + 1 $\label{l:binarySearch-bounding-inv-2}$
  @ loop_invariant high < toIndex $\label{l:binarySearch-bounding-inv-3}$
  @ loop_invariant !TArrays.has(a,fromIndex,low,key) $\label{l:binarySearch-essential-inv-1}$
  @ loop_invariant !TArrays.has(a,high+1,toIndex,key) $\label{l:binarySearch-essential-inv-2}$
  @*/
\end{lstlisting}
\caption{Loop invariants required for verifying method \<binarySearch0>.}
\label{fig:binary-search-inv}
\end{figure}

\subsection{Summary of the \DYNAMATE Approach}
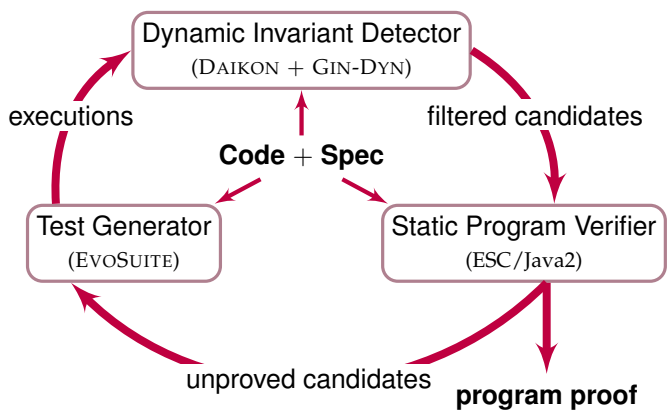
\begin{figure}[!b]
  \centering
\begin{adjustwidth}{-5mm}{-5mm}
  \begin{tikzpicture}[
  tool/.style={rectangle, minimum width=25mm,minimum height=10mm, very thick,rounded corners=2mm,draw=purple!50!black!50,font=\normalsize}, 
  node distance=15mm and -12mm,
  align=center
  ]
  \lstset{basicstyle=\footnotesize}

  \node (input) [color=black] {\textsf{\textbf{Code $+$ Spec}}};

  \node (invariants) [tool, above=6mm of input] {\textsf{Dynamic Invariant Detector} \\ {\footnotesize (\DAIKON $+$ \GINDYN)}};
  \node (prover) [tool,below right=of invariants] {\textsf{Static Program Verifier} \\ {\footnotesize (\ESCJ)}};
  \node (testing) [tool,below left=of invariants] {\textsf{Test Generator} \\ {\footnotesize (\EVOSUITE)}};

  \begin{scope}[color=purple, line width=3pt,every node/.style={fill=white},inner sep=2pt,outer sep=2pt,text=black]
]
  \draw (invariants.east) edge[-latex', bend left] node {\textsf{filtered candidates}} ($(prover.north) + (4mm,0)$);
  \draw ($(testing.north) + (-9mm,0)$) edge[-latex', bend left] node {\textsf{executions}} (invariants.west);
  \draw ($(prover.south)+(3mm,0)$) edge[-latex', bend left=45] node {\textsf{unproved candidates}} ($(testing.south)+(-7mm,0)$);

  \draw ($(prover.south)+(3mm,0)$) edge[-latex', line width=3pt]  node[below=6mm] {\textsf{\textbf{program proof}}} ++(0,-12mm);
  \end{scope}

  \begin{scope}[-latex', color=purple, ultra thick]
  \draw (input) edge (invariants.south);
  \draw (input) edge ($(prover.north) + (-18mm,0)$);
  \draw (input) edge ($(testing.north) + (+12mm,0)$);
  \end{scope}
\end{tikzpicture}
\end{adjustwidth}
\let\small=\footnotesize
    \caption{How \DYNAMATE works. The program code (center) is first fed into a test case generator (left), which generates executions covering legal behavior.  From these, two dynamic invariant detector tools (top) mine possible loop invariants, based both on fixed patterns (\DAIKON) as well as  postconditions (\GINDYN).  The candidates not invalidated by the generated runs are then fed (together with code and specification) into a symbolic program verifier (right).  The verifier then may produce a program proof (bottom right), but may also refute candidates, which initiates another round of executions, and thus refined invariants.}
    \label{fig:dynamate}
\end{figure}

In this paper, we present an approach that automates the functional verification of partial correctness of programs with loops by inferring the required loop invariants. 
The approach combines different techniques as illustrated in \autoref{fig:dynamate}.
Initially, a static verifier runs on the program code and its specification. 
If the verifier fails to prove the program correct, a round of four steps begins.

\nicepar{Step 1: test cases.}
To support dynamic invariant detection, a \emph{test case generator} builds executions of the program that satisfy the given precondition.

(In the \<binarySearch0> example, a possible test case searches for \<5> in the array \<[0, 1, 2]> and returns the index \<-3>.)

\nicepar{Step 2: candidate invariants.}
From the resulting executions, an invariant detector dynamically mines candidates for loop invariants.  Our \emph{invariant detector} tools use fixed, generic patterns as well as \emph{variable and specific patterns} derived from the given postcondition.  The patterns systematically determine candidates; only those candidates that hold for all executions are retained and go to the next stage.

(This step generates: $\mcode{high} < \mcode{toIndex}$ using the generic patterns; and \<!TArrays.has(a, high,>\linebreak \<toIndex, key)> from the postcondition. Neither is invalidated by the test case with \<a = [0, 1, 2]> and \<key = 5>, but the latter candidate does not hold in general.)

\nicepar{Step 3: invariant verification.}
The surviving set of loop invariant candidates are fed into a static \emph{program verifier}.  The verifier may confirm that some candidates are valid loop invariants.  

(The program verifier confirms that $\mcode{high} < \mcode{toIndex}$ is a valid loop invariant, but rejects the other candidate.)

\nicepar{Step 4: program verification and refinement.}
Using the verified invariants, the static verifier may also be able to produce a proof that the program is correct with respect to its specification.
If the proof does not succeed using the loop invariants inferred so far, \emph{another round} of generating, mining, and verifying starts.  If the verifier has left loop invariants unproved, the test generator searches for executions that falsify them, thus refining the set of candidates for the next proof attempt.

(A test case searching for \<2> in \<[0, 1, 2]> reveals that \<!TArrays.has(a, high, toIndex, key)> is invalid: it does not hold initially when $\mcode{high} = \mcode{toIndex} - \mcode{1}$.)
%

\nicepar{}
We implemented this approach in a tool called \DYNAMATE{}\footnote{\DYNAMATE = ``\underbar{Dyna}mic \underbar{M}ining \underbar{a}nd \underbar{Te}sting''}, a fully automatic verifier for \JAVA programs with loops.

\subsection{Loop Invariants from Postconditions}
A key functionality in \DYNAMATE is \GINDYN, a flexible mechanism for generating loop invariants.
While a vast amount of research on automatically finding loop invariants has been carried out over the years (\autoref{sec:related-work}), most approaches target restricted classes of loop invariants (such as linear inequalities over scalar numeric variables), which limits wide applicability in practice.
In fact, while  \ESCJ can handle full-fledged \JML-annotated \JAVA~programs, it provides no support for loop invariant discovery.

To overcome this limitation, in previous work~\cite{furia2010,LoopInvariantSurvey-TR-19112012} we introduced the idea of \emph{guessing invariant candidates based on mutations of the postcondition.}
The rationale for this idea is the observation that a loop drives the program state towards a goal characterized by the postcondition; a loop invariant, which characterizes all program states reached by the loop, can then be seen as a relaxation of the postcondition.
Mutating the postcondition is thus a way to obtain possible loop invariant candidates\iflong, which include assertions such as lines 26--27 in \autoref{fig:binary-search-inv}\fi.
The fundamental challenge in implementing this idea as a practically applicable technique is dealing with the blow-up in the number of candidates: generation is exponentially expensive and too many useless candidates bog down the program verifier with irrelevant or incorrect information.

In \autoref{sec:ginpink_approach}, we introduce the \GINDYN technique, which addresses this challenge.  \GINDYN uses dynamic checking to quickly weed out incorrect loop invariant candidates, and static checking to prune formally correct but redundant or irrelevant ones.
This way, the program verifier only has to deal with a small, manageable set of likely useful loop invariants, while the invariant generation process remains flexible as it is not restricted to simple fixed assertion patterns.

\subsection{Experimental Evaluation}
\iflong
\DYNAMATE builds on \EVOSUITE~\cite{fraser2011} as test case generator, \DAIKON~\cite{ErnstCGN2001:TSE} as dynamic invariant detector, and \ESCJ~\cite{ESCJava2} as program verifier; most of its power, however, comes from implementing \emph{\GINDYN}, a technique which systematically derives loop invariant candidates from postconditions.  
\else
\fi
We conducted a case study that applied \DYNAMATE to 28 methods from the \<java.util> classes in the Java standard library, including the \<binarySearch0> method.\footnote{As \autoref{sec:case-study-preparation} explains in detail, we occasionally modified the methods to remove features unsupported by \ESCJ---without changing the methods' input/output behavior.}
\DYNAMATE automatically discovered all loop invariants in \autoref{fig:binary-search-inv} given the code and specification in \autoref{fig:binary-search-code} and \autoref{fig:binary-search-spec}, resulting in fully automatic verification of the \<binarySearch0> method.  
Across the whole case study, \DYNAMATE discharged 97\% of the proof obligations of all the methods, resulting in full correctness proofs for 25 of the 28 methods.
As we show in \autoref{sec:comparison}, these results correspond to an over 20\% 
improvement over state-of-the-art tools for automatic verification, such as the CodeContracts static checker, in terms of number of automatically discharged obligations.

Being able to provide such fully automatic proofs of real \JAVA code is certainly promising.  However, we have to adjust our expectations: from a software engineering perspective, the methods \DYNAMATE can prove are still very small.
Unfortunately, the state of the art is such that proofs of large, complex systems are simply impossible without significant manual effort by highly-trained experts; and the intrinsic complexity of formal verification indicates that they will remain so for the foreseeable future.
However, \emph{if} a critical function is similar in complexity to those analyzed in our experiments---that is, mostly array-based algorithm implementations with a handful of loops and conditions---then there are  chances that \DYNAMATE will be able to prove it. 
(And, if the proof attempt fails, no human effort is wasted.)  From a verification perspective, being able to conduct such proofs automatically on practical functions from production code is an important goal; \DYNAMATE neither expects nor requires any loop annotation, user interaction, or expert knowledge.

\subsection{Summary of Contributions}
The main contributions of this paper are:

\begin{enumerate} 
\item \DYNAMATE: an algorithm to automatically discharge proof obligations for programs with loops, based on a combination of dynamic 
and static 
techniques.
\item \GINDYN: an automatic technique to boost the dynamic detection of loop invariants, based on the idea of syntactically mutating postconditions~\cite{furia2010}. 
\item A prototype implementation of the \DYNAMATE algorithm that integrates the \EVOSUITE test case generator, the \DAIKON dynamic invariant detector, and the \ESCJ static verifier, as well as \GINDYN. 
\item An evaluation of our \DYNAMATE prototype on a case study involving 28 methods with loops from \<java.util> classes. 
\item A comparison against state-of-the-art tools for automatic verification based on predicate abstraction, abstract interpretation, and con\-straint-based techniques.
\end{enumerate}

 The remainder of this paper is organized as follows.
 \autoref{sec:dynamate} overviews the \DYNAMATE algorithm in general, followed by a description of the current prototype in \autoref{sec:loopmateaction}.  
 \autoref{sec:ginpink_approach} describes in more detail the \GINDYN technique for loop invariant detection.
 \autoref{sec:evaluation} evaluates \DYNAMATE on a case study consisting of \JAVA code from selected \<java.util> classes.  \autoref{sec:related-work} discusses related work in loop invariant detection.  \autoref{sec:conclusion} closes with conclusions and future work.

\section{Overview of the DYNAMATE Algorithm}
\label{sec:dynamate}

\DYNAMATE builds on an interplay of three components: a test case generator, a dynamic invariant detector, and a static verifier.  
The algorithm is applicable to any language that offers three such components.

\label{sec:loopmate_algo}

The \DYNAMATE algorithm, illustrated in Algorithm~\ref{alg:loopmate}, inputs a program $M$ and its specification---a precondition $P$ and a postcondition $Q$. 
Two outcomes of the algorithm are possible: \emph{success} means that \DYNAMATE has found a set of valid loop invariants that are sufficient to statically verify $M$ against its specification $(P, Q)$; \emph{failure} means that \DYNAMATE cannot find new valid loop invariants, and those found are insufficient for static verification.
Even in case of failure, the valid loop invariants found by \DYNAMATE have a chance of enabling \emph{partial} verification by discharging some proof obligations necessary for a correctness proof.

\begin{algorithm}
\caption{The \DYNAMATE algorithm}
\begin{algorithmic} 
\label{alg:loopmate}
\REQUIRE Program $M$, precondition $P$, postcondition $Q$ 
\STATE $\|TS| \leftarrow \emptyset$       $\qquad\;$ (set of tests)
\STATE $\|INV| \leftarrow \emptyset$      $\qquad$(set of verified loop invariants)
\STATE $C$ $\leftarrow  \emptyset$         $\qquad$(set of candidates)
\WHILE {static verifier cannot prove $(M,P,Q, \|INV|)$}
	\STATE $\|T| \leftarrow$ execute test case generator on $(M,P,C)$
	\STATE $\|TS| \leftarrow \|TS| \cup \|T|$
	\STATE $I$ $\leftarrow$ execute invariant detector on $(M,\|TS|)$
	\IF{$I$ has not changed}
		\STATE return (``failure'', $\|INV|$ )
	\ENDIF
	\STATE $\widehat{M}\leftarrow$ annotate $M$ with candidate invariants $I$
	\STATE $J$ $\leftarrow$ statically check valid invariants of $(\widehat{M},P)$
   \STATE $\|INV| \leftarrow  \|INV| \:\cup\: J$
   \STATE  $C \leftarrow I \setminus\|INV|$
\ENDWHILE 
\STATE return (``success'', $\|INV|$)
\end{algorithmic}
\end{algorithm}

\DYNAMATE's main loop starts by executing the test case generator, which produces a new set \|T| of test cases that exercise $M$ with inputs satisfying the precondition $P$.
The loop feeds the overall set \|TS| of test cases generated so far to the dynamic invariant detector, which outputs a set of candidate loop invariants $I$.
We call them ``candidates'' because the invariant detector summarizes a finite set of runs, and hence it may report assertions that are not valid loop invariants in general.

To find out which candidates are indeed valid, \DYNAMATE calls the static verifier on the program annotated with all candidates $I$; 
the verifier returns a set of proved candidates $J$ (a subset of $I$), which \DYNAMATE adds to the set \|INV| of verified loop invariants.

Then, using the current \|INV|, it calls the static verifier again, this time trying a full correctness proof of $M$ against $(P, Q)$.
If verification succeeds, \DYNAMATE terminates with success.
Otherwise, \DYNAMATE's main loop continues as long as the invariant detector is able to find new candidate invariants, to invalidate previous candidates, or both.
The loop may also diverge, keeping on finding new candidate invariants until it runs out of resources.

Since we assume a static verifier that is sound but incomplete, unproved candidates in $I \setminus \|INV|$ are not necessarily invalid.
Thus, in case of failed proof, \DYNAMATE relies on the test case generator to create executions of $M$ that invalidate as many unproved candidates in $C$ as possible.

\section{How DYNAMATE Works}
\label{sec:loopmateaction}

This section details how \DYNAMATE works, using \linebreak\<binarySearch0> as running example.

\subsection{Input: Programs and Specifications}
\label{sec:input}

Each run of \DYNAMATE takes as input a \JAVA method $M$ with its functional specification consisting of precondition $P$ and postcondition $Q$.
Pre- and postcondition are written in \JML~\cite{leavens2006}: an extension of the \JAVA syntax for expressions that supports standard logic constructs such as implications and quantifiers. 
$P$ and $Q$ generally consist of a number of \emph{clauses}, each denoted by the keyword \<requires> (precondition) and \<ensures> (postcondition); clauses are implicitly logically conjoined.
In \autoref{fig:binary-search-spec}, for example, the precondition has three clauses and the postcondition has two.

While \DYNAMATE can work with \JML specifications in any form, we find it effective to follow the principles of the model-based approach to specification (introduced with \JML~\cite{leavens2006} and developed in related work of ours~\cite{PFM10-VSTTE10,PFPWM-ICSE13}).
Specifications refer to underlying mathematical models of the data structures involved; a collection of predicates encapsulate the assertions relevant to the application domain and express them abstractly in terms of the underlying models.
The specification of \<binarySearch0> in \autoref{fig:binary-search-spec} follows this approach, as it uses the predicates \<within>, \<sorted>, and \<has>, which are collected in a class \<TArrays> providing notation to capture facts about integer array-like data structures.

Following the model-based specification style entails three main advantages for our work.
First, it improves the abstraction and clarity of specifications, and hence it also facilitates reuse with different implementations.
For instance, it should be clear that \<has(a, fromIndex, toIndex, key)> means that array \<a> contains a value \<key> within \<fromIndex> and \<toIndex>.
Using predicate \<has> also abstracts several details about how arrays are implemented, such as whether they are zero-indexed; in fact, we would write the same specification even if \<a> were, say, a dynamic list and \<fromIndex> and \<toIndex> two pointers to list elements.
Predicates are defined only once and then reused for all methods working in a similar domain; in fact, for our experiments (\autoref{sec:evaluation}) we developed a small set of predicates in two classes (\<TArrays> and \<TLists>), which were sufficient to express the specification and loop invariants of all 28 methods used in the evaluation.

Second, model-based specifications also make it easy to reconcile static and a runtime semantics. 
When developing predicates in \<TArrays> we defined each predicate as a \<static boolean> method with both a \JAVA implementation and a \JML specification.
For example, \<has>'s precondition declares its domain of definition (in words: \<a> is not \<null> and \<[fromIndex..toIndex)> is a valid interval within \<a>'s bounds); its postcondition
\begin{center}
\begin{lstlisting}[basicstyle=\normalsize\ttfamily]
\exists int i; fromIndex <= i && i < toIndex && key == a[i]
\end{lstlisting}
\end{center}
defines the static semantics of the predicate using explicit quantification and primitive operations; its implementation recursively searches through \<a> for an element \<key> and returns \<true> iff it finds it within \<fromIndex> and \<toIndex>.
Dynamic tools will then use its implementation to check whether a predicate holds; static tools will instead understand a predicate's semantics in terms of its \JML specification.

A third advantage of using model-based specifications is leveraged by the \DYNAMATE approach and more precisely by the \GINDYN invariant detector described in \autoref{sec:ginpink}.
Collections of predicates such as \<TArrays> provide useful domain knowledge, as they suggest which predicates are likely to be pertinent to the specification of the methods at hand: if a method's postcondition $Q$ includes a predicate $p$ from some class $T$, it's likely that the loop invariants necessary to prove $Q$ also include $p$ or some other predicates defined in $T$.

\subsection{Test Case Generation}
\label{sec:evosuite}

The \DYNAMATE algorithm needs concrete executions to dynamically infer loop invariants: \DAIKON mines relations that hold in all passing test cases (\autoref{sec:daikon}); and \GINDYN filters out invalid loop invariant candidates that are falsified by a test case (\autoref{sec:ginpink}). 
While any test case generator could work with \DYNAMATE, our prototype integrates \EVOSUITE~\cite{fraser2011}, a fully automatic search-based tool using a genetic algorithm.
Besides being a fully automated tool, a specific advantage of \EVOSUITE is that its genetic algorithm evolves test suites towards covering all program branches at the same time, and hence infeasible branch conditions (common in the presence of assertions) do not ultimately limit search effectiveness.

Within \DYNAMATE, \EVOSUITE needs to generate test cases that satisfy preconditions written in \JML.
To this end, we instrument every method with an initial check that its precondition holds: if it does not, the instrumentation code throws an exception that terminates execution and makes the test case invalid.
The instrumentation leverages the availability of a runtime semantics for all specification pred\-i\-cates---developed using the model-based style discussed in \autoref{sec:input}.

One of \<binarySearch0>'s preconditions requires that the input array \<a> be sorted within the range from \<fromIndex> to \<toIndex>; \autoref{fig:runtime-predicate} shows the corresponding instrumentation that calls the implementation of the predicate \<sorted> and throws an exception if it evaluates to false.

Since \EVOSUITE tries to maximize branch coverage, it has a good chance\footnote{Due to randomization, success may vary between runs.} of producing tests that pass all precondition checks and thus represent valid executions according to the specification.
For example, \EVOSUITE produces the valid test in \autoref{fig:binary-search-tests} where the value $-1030$ is searched for in a trivially sorted array initialized to all zeros.

\begin{figure}[t]
\begin{lstlisting}
private static int binarySearch0(int[] a, 
       int fromIndex, int toIndex, 
       int key) {
  // Precondition clause: 
  // #TArrays.sorted(a,fromIndex,toIndex)#
  if (!TArrays.sorted(a,fromIndex,toIndex))
    throw new AssertionError("Failed precondition");
  ...
\end{lstlisting}
\caption{Runtime check for a precondition clause of \<binarySearch0>'s.}
\label{fig:runtime-predicate}
\end{figure}

\begin{figure}[t]
\begin{lstlisting}
void test() {  
 // Element not found
 int[] intArray = new int[8];
 int intV = Arrays.binarySearch0(intArray, 0, 6, -1030);
}
\end{lstlisting} 
\caption{A test that covers the ``not found'' branch in \<binarySearch0>.}
\label{fig:binary-search-tests}
\end{figure}

\subsection{Dynamic Loop Invariant Inference}
\label{sec:daikon}

At the core of the \DYNAMATE algorithm lies a component that detects ``likely'' loop invariants based on the concrete executions provided by the test case generator.
The current \DYNAMATE implementation relies on two modules with complementary functionalities\iflong to infer loop invariants dynamically\fi.
Simple boilerplate invariants (mostly involving numeric relations between sca\-lar variables or basic array properties) come from \DAIKON~\cite{ErnstCGN2001:TSE}.
On top of it,  \GINDYN produces more complex invariants---involving the same predicates used in the specification---discovered by mutating postcondition clauses.

\DAIKON's and \GINDYN's invariants are complementary; for example, neither one suffices for a correctness proof of  \<binarySearch0>.
\iflong On the other hand, \fi\DAIKON invariants are usually necessary as a basis to establish \GINDYN invariants (for example, to prove a predicate is well-defined by ensuring absence of out-of-bound errors).
Therefore, we run \DAIKON first and introduce \GINDYN invariants as soon as we reach a fixpoint in the overall \DYNAMATE algorithm.
This scheme is flexible and fosters an effective combination between loop invariant inference techniques with complementary advantages.
The rest of this section describes how \DYNAMATE uses \GINDYN  and \DAIKON.

\subsubsection{Dynamic Invariant Detection with {\DAIKON}}

\DAIKON~\cite{ErnstCGN2001:TSE} is a widely used dynamic invariant detector 
which supports a set of basic invariant templates.
Given a test suite and a collection of program locations as input, \DAIKON instantiates its templates with program variables, and traces their values at the locations in all executions of the tests.
The instantiated templates that hold in every execution are retained as likely invariants at those locations.
Likely invariants are still just candidates (they may be invalid) because they are based on a finite number of executions.

\begin{table}
\centering
\caption{\textnormal{Some loop invariant candidates produced by \DAIKON in the first iteration of \textsc{dynamate}. Column \textsc{valid} reports which candidates are valid.}}
\label{table:daikon_inv_1}
\begin{tabular}{rlc}
\toprule
\textsc{id} & \textsc{candidate} & \textsc{valid} \\
\midrule
$c_1$ & \<key> $\in$ $\{-1030, 0\}$ & \textsc{no}  \\
$c_2$ & \<a> $\neq$ \<null> & \textsc{yes} \\
$c_3$ & \<a[]>'s elements one of $\{-915,0\}$ & \textsc{no} \\
$c_4$ & \<Arrays.INSERTION\_THRESHOLD> $\neq$ \<toIndex> & \textsc{no} \\
$c_5$ & \<low> $\geq$ \<fromIndex> & \textsc{yes}\\
$c_6$ & \<low> $\geq$ \<key> & \textsc{no} \\
$c_7$ & \<high> $<$ \<toIndex> & \textsc{yes}\\
$c_8$ & \<high> $>$ \<key> & \textsc{no} \\
$c_9$ & \<high> $\leq$ \<a.length - 1> & \textsc{yes}\\
$c_{10}$ & \<toIndex> $>$ \<fromIndex> & \textsc{no} \\
\bottomrule
\end{tabular}
\end{table}

Since \DYNAMATE needs \emph{loop} invariants, it instructs \DAIKON to trace variables at four different locations of each loop: before loop entry, at loop entry, at loop exit, and after loop exit.
To this end, we instrument the input programs adding dummy methods invoked at these locations, which gives access to all variables within the loops without requiring modifications to \DAIKON.

\DYNAMATE executes \DAIKON on \<binarySearch0> using the initial test suite, which produces 62~loop invariant candidates, most of which are invalid due to the incompleteness of the initial test suite.
\autoref{table:daikon_inv_1} shows 10 of the 62 candidates; among others, $c_3$ is merely a reflection of the fact that the initial test suite only put the values \<0> and $-$\<915> in the arrays.
Another visible characteristic of these invariants is that they are limited to simple properties, and hence they are insufficient to prove \<binarySearch0>'s postcondition.
In general, \DAIKON templates give invariants that are immediately useful to establish simple properties such as absence of out-of-bound accesses, but are hardly sufficient to prove full functional correctness.

\subsubsection{{\GINDYN}: Invariants from Postconditions}
When the \DYNAMATE algorithm reaches a fixpoint without finding a proof, it is a sign that \DAIKON has run out of steam and more complex invariants are required to make progress.
For example, \<binarySearch0>'s loop invariants on lines~\ref{l:binarySearch-essential-inv-1}--\ref{l:binarySearch-essential-inv-2} in \autoref{fig:binary-search-inv} are never produced by \DAIKON with its standard templates.

Extending \DAIKON with new templates is possible, but even leaving implementation issues aside (e.g., \DAIKON only includes non-static methods with no arguments as predicates) we cannot anticipate all possible forms a loop invariant may take.
Instead, we apply ideas introduced in our previous work~\cite{furia2010}: since loop invariants can be seen as relaxations of the postcondition, syntactically mutating a postcondition $Q$ generates variants of $Q$ which we use as suggestions for invariant \emph{candidates}.
We turn this intuition into a practically applicable technique by providing \GINDYN: a way to efficiently generate and filter out a large amount of invalid or uninteresting invariant candidates.
Section~\ref{sec:ginpink} describes in detail how \GINDYN  does the filtering, again based on a combination of dynamic and static techniques.
The rest of the current section briefly discusses how invariant candidates produced by  \GINDYN are used within \DYNAMATE.

When it reaches a fixpoint,  \DYNAMATE  asks \GINDYN  for a new \emph{wave} of invariant candidates, adds them to the set of current candidates, and proceeds with static validation, which is possibly followed by other iterations of the main algorithm.
Thus, \GINDYN and \DAIKON invariant candidates are used in the very same way; the fact that they nearly always are disjoint sets increases the chance of eventually getting to a complete set of invariants.

In fact, \GINDYN produces the two fundamental invariants on lines~\ref{l:binarySearch-essential-inv-1}--\ref{l:binarySearch-essential-inv-2} in \autoref{fig:binary-search-inv} necessary for a correctness proof of \<binarySearch0>.
The final set of \emph{verified} loop invariants includes those of \autoref{fig:binary-search-inv} with 28 more,\footnote{In this work, we do not address minimizing the number of invariants.} consisting of $13$ invariants found by \DAIKON and $20$ invariants found by \GINDYN.

\subsection{Static Program Verification}
\label{sec:vcc}

In \DYNAMATE, invariant candidates come from dynamic analysis; hence they are only educated guesses  based on a finite number of tests.
The \DYNAMATE algorithm complements dynamic analysis with a \emph{static} program verifier, which serves two purposes: (1) verifying loop invariant candidates, and (2) using verified loop invariants to carry out a conclusive correctness proof.

\subsubsection{Verification of Loop Invariants}
\label{subsec:validationofloopinvariants}

The \DYNAMATE prototype relies on the \ESCJ~\cite{ESCJava2} static verifier, which works on \JAVA programs and \JML annotations.

By default~\cite{Kiniry:2006:SCW:1181195.1181200}, \ESCJ handles loops unsoundly by unrolling them a finite number of times; but, by enabling the \texttt{-loopSafe} option, it encodes the exact modular semantics of loops based on loop invariants~\cite[Sec.~5.2.3]{Burdy:2005:OJT:1070908.1070911}.\footnote{See also: \url{http://sourceforge.net/p/jmlspecs/mailman/message/31579805/}}
Therefore, \DYNAMATE always calls \ESCJ with the \texttt{-loopSafe} option enabled.

To determine whether a candidate $L$ is a valid invariant of some loop $\ell$, \DYNAMATE annotates $\ell$ with the assertion $\mcode{loop_invariant} \;L$ and runs \ESCJ on the annotated program.
As part of verifying the input, \ESCJ checks whether $L$ is a valid loop invariant, that is it holds initially and is preserved by every iteration.
\DYNAMATE searches for \ESCJ \emph{warnings} that signal whether checking $L$ was successful.
Since \ESCJ is a sound but incomplete tool,\footnote{On the Java code we apply it to: see Footnote~\ref{footnote:note-soundness}} if it produces no warning about~$L$, we conclude that $L$ is a valid loop invariant; if it does produce a warning, it may still be that $L$ is valid but the current information (typically, in the form of other loop invariants) is insufficient to establish it with certainty.

Due to the undecidability of first-order logic, \ESCJ may also time out or run out of memory, which also counts as inconclusive output.
In all such cases, the \DYNAMATE algorithm leverages the static/dynamic complementarity once again: it initiates another iteration of testing, trying to conclusively falsify the candidates that \ESCJ could not validate.

Invariant candidates to be validated may have dependencies, that is proving that one candidate $L_1$ is valid requires to assume another candidate $L_2$.
To find a maximal set of valid loop invariants, \DYNAMATE applies the \HOUDINI algorithm~\cite{flanagan2001}, which first considers all candidates at once, and then iteratively removes those that cannot be verified until all surviving candidates are valid loop invariants.
In the \<binarySearch0> example, \DAIKON produces 62 loop invariant candidates in the first iteration.
Running \ESCJ determines that half of them are valid loop invariants, among which $c_2$, $c_5$, $c_7$, and $c_9$ in \autoref{table:daikon_inv_1}.

\subsubsection{Program Proof}
At the end of each iteration, \DYNAMATE uses the current set of valid loop invariants to attempt a correctness proof of the program against its specification.
If \ESCJ succeeds, the whole \DYNAMATE algorithm stops with success; otherwise, it begins another iteration with the goal of improving its invariant set.

The first iteration of \DYNAMATE on \<binarySearch0> does not find a correctness proof, since \iflong we already mentioned that \fi the invariants in \autoref{table:daikon_inv_1} are insufficient to prove the specification in \autoref{fig:binary-search-spec}.

\subsection{Refining the Search for Loop Invariants}
Unproved loop invariant candidates may be over-specific and hence unsound---such as $c_1$ in \autoref{table:daikon_inv_1}. 
Since this may indicate unexplored program behavior, for every such candidate $L$, \DYNAMATE adds the conditional check  
\begin{lstlisting}[mathescape]
    if ($\neg L$) throw new AssertionError("$L$ failed")
\end{lstlisting}
at the locations where loop invariants are checked.
The check is enclosed by a \<try/catch> block, so that testing $L$ does not affect testing other candidates.
This directs \EVOSUITE's search towards trying to explore the new conditional branch, and hence falsifying $L$.
This helps refine the dynamic detection of loop invariants by providing more accurate tests.

\section{GIN-DYN: Loop In\-vari\-ants from \\ Postconditions}
\label{sec:ginpink_approach}
\label{sec:ginpink}

The postcondition $Q$ of a method $M$ characterizes the goal state of $M$'s computation; a loop $\ell$ in $M$'s body contributes to reaching the goal by going through a series of intermediate states, which $\ell$'s loop invariants characterize.
Based on these observations~\cite{LoopInvariantSurvey-TR-19112012}, we suggested~\cite{furia2010} the idea of systematically \emph{mutating} $Q$ to guess candidate invariants of $\ell$.

The throwaway prototype discussed in~\cite{furia2010} did not work on a real programming language, and required users to manually select a subset of mutation operators to direct the generation of candidates.
In fact, applying these ideas in a fully automatic setting on realistic programs requires to address two fundamental challenges:
(a)~providing a sweeping widely-applicable collection of mutations, and 
(b)~efficiently discarding a huge number of invalid and uninteresting mutations.

The \GINDYN technique presented in this section ad\-dress\-es these challenges by following a three-step strategy:
\begin{enumerate}
\item Generate many \emph{mutants} (i.e., variants) of $Q$'s clauses, applying predefined generic sequences of mutation operators. 
\item Discard mutants that cannot be loop invariants of $\ell$ because at least one test case exercising $\ell$ violates them. 
\item Further prune valid but uninteresting loop invariants by eliminating those mutants that are \emph{tautologies}, that is that hold independent of the specific behavior of $\ell$.
\end{enumerate}
The mutants that survive both the validation and the tautology elimination phases are in a good position to be useful to establish $Q$ by virtue of having been derived from it.
The following subsections detail the three steps as they are available in \DYNAMATE.

\subsection{Generation of Mutants}
The fundamental idea behind \GINDYN is to syntactically mutate postcondition clauses by substituting an expression for another one.
The mutation algorithm maintains a set $\muts$ of \emph{mutants} and, for each type \<t>,  a set $\expr{\<t>}$ of available expressions of type \<t>.
$\muts$ initially consists of the clauses in postcondition $Q$, and $\expr{\<int>}$ and $\expr{\<int[]>}$ include the integer and integer array variables available in the current loop $\ell$.
The basic mutation operators 
(or just ``mutations'' for short) 
used by \GINDYN are the following:
\begin{itemize}
\item \textbf{Substitution:} given a mutant $m \in \muts$ and an expression $e \in \expr{\<t>}$ of type \<t>, replace by $e$ any subexpression of type \<t> in $m$. This generates as many mutants as are subexpressions of type \<t> in $m$, which are added to $\muts$.
\item \textbf{Weakening:} given a mutant $m \in \muts$ and a Boolean expression $b \in \expr{\<boolean>}$, 
add to $\muts$ the mutants $b \limpl m$ and $\lnot b \limpl m$.
\item \textbf{Aging:} given a mutant $m \in \muts$, replace by $e - 1$ or $e + 1$ any subexpression $e$ of type \<int> in $m$. This generates twice as many mutants as are subexpressions of type \<int> in $m$, which are added to $\muts$.
\end{itemize}
In addition to these operators, the mutation algorithm can extend the sets of available expressions, thus enabling a larger number of substitutions; in particular, \textbf{predicate extraction} is the process of adding to the set $\expr{\<boolean>}$ of Boolean expressions all predicates belonging to the same collection (such as \<TArrays>) as those in the postcondition (see Section~\ref{sec:input}).

\begin{figure}[ht!]
\begin{lstlisting}[mathescape]
$Q$: 
 \result $<$ 0 ==> $\;\neg$TArrays.has(a,fromIndex,toIndex,key)
$\leadsto$ low $<$ 0 ==> $\;\neg$TArrays.has(a,fromIndex,toIndex,key) 
$\leadsto$ low $<$ 0 ==> $\;\neg$TArrays.has(a,fromIndex,mid,key)
\end{lstlisting}
\caption{Mutations producing a valid but trivial loop invariant.}
\label{fig:mutations-useless}
\end{figure}

\begin{figure}[t]
\scriptsize
\begin{lstlisting}[mathescape]
$Q$: 
 \result $<$ 0 ==> $\;\neg$TArrays.has(a,fromIndex,toIndex,key) 
$\leadsto$  $\neg$TArrays.has(a,fromIndex,toIndex,key)
$\leadsto$  $\neg$TArrays.has(a,high,toIndex,key)
$\leadsto$  $\neg$TArrays.has(a,high + $\,$1,toIndex,key)
\end{lstlisting}
\caption{Mutations producing a valid and useful loop invariant.}
\label{fig:mutations-useful}
\end{figure}

\autoref{fig:mutations-useless} shows one example of applying two substitutions to \<binarySearch0>'s second postcondition clause: \<\\result> is replaced with \<low>; and  \<toIndex> is replaced with \<mid>.
The produced mutant is clearly not a useful loop invariant because its antecedent is always false in the loop's context.
In contrast, the mutations applied in \autoref{fig:mutations-useful} produce one of \<binarySearch0>'s required loop invariants: predicate extraction adds \<!has(a, fromIndex, toIndex, key)> to the set of available Boolean expressions; a first substitution uses it to replace the whole postcondition; one further \<int> substitution of \<high> for \<fromIndex> and one application of aging (turning \<high> into \<high + 1>) determine a valid loop invariant which is also useful in the correctness proof.

\subsubsection{Mutation Waves}

The number of combined multiple substitutions dominates the combinatorial complexity of generating mutants: if we start with $q$ postcondition clauses in $\muts$, each having up to $s$ subexpressions of  some type \<t>, and we have $e$ expressions of type \<t> available, applying all possible $n$ consecutive \<t>-type substitutions builds a number of mutants in the order of $(qse)^n$.
Therefore, for all but the most trivial cases it is unfeasible to exhaustively apply more than few multiple substitutions.

To mitigate this problem, the mutation algorithm combines multiple applications of mutation operators (possibly with predicate extraction) to determine complex mutants that may significantly differ from the initial postcondition\iflong but are still derived from it\fi.
\GINDYN defines several mutation sequences, which we call \emph{waves}; each wave $\omega$ combines a variable number of mutations, which are applied exhaustively starting with the postconditions and finally producing a set $\muts_\omega$ of mutants.

\input{waves.tex}

\subsection{Validation of Mutations}

Each wave $\omega$ produces a set $\muts_\omega$ of mutants.
Mutants are educated guesses; the large majority of them are not loop invariants of $\ell$ and must be discarded.
To determine whether a mutant $\mu \in \muts_\omega$ is a loop invariant, \GINDYN injects a check of $\mu$ at the entry and exit of $\ell$'s body, runs the test cases available for $\ell$ (generated as in \autoref{sec:evosuite}), and retains in $\muts_\omega$ only the mutants that pass all tests.

Being based on dynamic techniques, the validation of mutations performed within  \GINDYN is provisional, in that validated mutants still have to pass muster with the static program prover before we can conclude they are valid loop invariants.
However, one advantage of testing is that we can check mutants for invariance \emph{independently} of one another.
In contrast, static verifiers such as \ESCJ work modularly: they reason about the program state at the beginning of every loop iteration entirely in terms of the available loop invariants, and hence they may fail to confirm that a given assertion $L$ is indeed a loop invariant if other invariants, necessary to prove $L$, are not provided.
The dependency problem is particularly severe in the presence of nested loops, where there can be circular dependencies. 
Runtime checks do not incur such limitations, and can check mutants for invariance in batches; the batch size is determined by how many assertions we can monitor at once, whereas how we partition the mutants in batches does not affect the correctness of the final result.

\subsection{Tautology Elimination}
Several of the mutants $\muts_\omega$ that pass validation are trivial,
 in that they hold only by virtue of their logic structure---that is, they are tautologies, useless to prove the postcondition~$Q$.
For example, many mutants are implications with an identically false antecedent (such as the one in \autoref{fig:mutations-useless}).
To identify and discard mutants that are tautologies, \GINDYN uses \ESCJ as follows.
For each mutant $\mu \in \muts_\omega$, it builds a dummy method $m_\mu$ with the following structure:
\begin{lstlisting}[mathescape]
   public void $m_\mu$(t$_1$ v$_1$, t$_2$ v$_2$, $\ldots$)  {
     //@ assume $\textrm{\textit{verified loop invariants}}$
     //@ assert $\mu$
   }
\end{lstlisting}
The arguments of $m_\mu$ include all variables occurring in $\mu$; since \ESCJ reasons modularly, it makes no assumption about their values, which is tantamount to setting them nondeterministically.
Then, an \<assume> lists all loop invariants verified across all \DYNAMATE iterations by \ESCJ (\autoref{subsec:validationofloopinvariants}).
Any formula that is a consequence of these already known invariants is redundant and should not be retained.
The final \<assert> asks \ESCJ to establish the mutant $\mu$ in the given context; \GINDYN retains $\mu$ only if \ESCJ \emph{cannot} prove the \<assert>, and hence $\mu$ is not logically derivable from the verified loop invariants.
This way, the tautology elimination is sound and complete relative to the static verifier's capabilities.

\section{Case Study}
\label{sec:evaluation}

\begin{table*}[t]
\caption{%
\textnormal{%
Signatures of methods in \<java.util> selected as a basis for our case study. We show the number of lines (\textsc{loc}) of each method's body, the number of loops ($|\ell|$, where $m \{n\}$ denotes $m$ outer loops and $n$ inner loops), and which language features were modified to obtain the corresponding methods of the case study in \autoref{table:stats-casestudy}.
}
}
\label{table:orig-casestudy}
\centering
\scriptsize
\setlength{\tabcolsep}{1.5pt}
\begin{tabular}{@{}l l r @{$\quad$} r@{$\;$}l@{} @{$\ $}l} 
\toprule
\textsc{class}&\textsc{method} & \textsc{loc}  & \multicolumn{2}{c}{$|\ell|$}
 & \multicolumn{1}{c}{\textsc{modified}}
\\
\midrule
\!ArrayDeque! & \!contains(Object)!                   &  11 &       1 &             
& bitwise mask 
\\
\!ArrayDeque! & \!removeFirstOccurrence(Object)!      &  13 &   1 &             
& bitwise mask
\\
\!ArrayDeque! & \!removeLastOccurrence(Object)!       &  13       &   1 &             
& bitwise mask, conditional structure 
\\
\!ArrayList! & \!clear()!                       &   4 &   1 &             
\\
\!ArrayList! & \!indexOf(Object)!                     &  10  &   2 &             
\\
\!ArrayList! & \!lastIndexOf(Object)!                 &  10 &   2 &             
\\
\!ArrayList! & \!remove(Object)!                      &  14 &   2 &             
& refactored calls to \!fastRemove!
\\
\!Arrays! & \!binarySearch0(int[],int,int,int)!                  &  13 &   1 &             
& bitwise mask
\\
\!Arrays! & \!equals(int[],int[])!                         &  11 &   1 &             
\\
\!Arrays! & \!fill(int[],int)!             &   2 &   1 &          
\\
\!Arrays! & \!fill(int[],int,int,int)!   &   3  &   1 &          
\\
\!Arrays! & \!fill(Object[],Object)!        &   2 &   1   &          
\\
\!Arrays! & \!fill(Object[],int,int,Object)!  &   3 &   1      &       
\\
\!Arrays! & \!hashCode(int[])!         &  6 &      1 &          
& ``for each'' loop expressed as regular \!for!
\\
\!Arrays! & \!hashCode(Object[])!   &  6 &   1    &       
& ``for each'' loop expressed as regular \!for!
\\
\!Arrays! & \!sort1(int[],int,int)! &  42 &   5  & $\{3\}$     
& loops factored out into \!insertionSort_b! and \!quicksortPartition!
\\
\!Arrays! & \!mergeSort(Object[],Object[],int,int,int)! &  25 &   3    & $\{1\}$             
& loops factored out into \!insertionSort_a! and \!merge!, for \!Integer!s
\\
\!Arrays! & \!vecswap(int[],int,int,int)!                        &  2  &   1    &             
& refactored local variables
\\
\!Collections! & $\text{\!replaceAll(List!}\langle\text{\!T!}\rangle\hspace{-1pt}\text{\!,T,T)!}$   &  37 &   4   &             
& retained only one replacement loop that uses no external iterators
\\
\!Collections! & $\text{\!reverse(List!}\langle\text{\!T!}\rangle\hspace{-1pt}\text{\!)!}$ &  13 &   2  &             
& retained only one replacement loop that uses no external iterators
\\
\!Collections! & $\text{\!sort(List!}\langle\text{\!T!}\rangle\hspace{-1pt}\text{\!)!}$ &   7 &   1  &             
& replaced external iterators with direct access
\\
\!Vector! & \!indexOf(Object,int)!                        &  10 &   2  &             
\\
\!Vector! & \!lastIndexOf(Object,int)!                    &  12 &   2 &             
\\
\!Vector! & \!removeAllElements()!              &   4 &  1 &             
\\
\!Vector! & \!removeRange(int,int)!                    &  7 &    1 &             
\\
\!Vector! & \!setSize(int)!                        &  9 &   1 &             
\\
\midrule
& \textsc{total}                              & 289   &  41 &  $\{4\}$ \\
\bottomrule
\end{tabular}
\end{table*}

\begin{table*}[t]
\caption{%
\textnormal{%
Annotated methods used as case study, obtained from those in \autoref{table:orig-casestudy}. We show the number of lines (\textsc{loc}) of each method's body, the number of pre- ($|P|$) and postcondition ($|Q|$) clauses given as specification, and the number of loops ($|\ell|$, where $m \{n\}$ denotes $m$ outer loops and $n$ inner loops).
}
}
\label{table:stats-casestudy}
\centering
\setlength{\tabcolsep}{1.5pt}
\begin{tabular}{@{}r l r r r @{$\quad$} r@{$\;$}l@{}} 
\toprule
\textsc{class}&\textsc{method} & \textsc{loc}  & $|P|$ & $|Q|$ & \multicolumn{2}{c}{$|\ell|$} 
\\
\midrule
\<ArrayDeque> & \<contains>                   &  13 &   0   &   2   &   1 &             
\\
\<ArrayDeque> & \<removeFirstOccurrence>      &  15 &   0   &   2   &   1 &             
\\
\<ArrayDeque> & \<removeLastOccurrence>       &  15 &   0   &   2   &   3 &             
\\
\<ArrayList> & \<clear>                       &   9 &   0   &   3   &   1 &             
\\
\<ArrayList> & \<indexOf>                     &  12 &   0   &   3   &   2 &             
\\
\<ArrayList> & \<lastIndexOf>                 &  12 &   0   &   3   &   2 &             
\\
\<ArrayList> & \<remove>                      &  14 &   0   &   5   &   2 &             
\\
\<Arrays> & \<binarySearch0>                  &  18 &   2   &   2   &   1 &             
\\
\<Arrays> & \<equals>                         &  16 &   0   &   1   &   1 &             
\\
\<Arrays> & \<fill\_a> (\<int> array)            &   4 &   1   &   1   &   1 &          
\\
\<Arrays> & \<fill\_b> (\<int> array range)      &   5 &   2   &   3   &   1 &          
\\
\<Arrays> & \<fill\_c> (\<Object> array)        &   4 &   1   &   3   &   1 &          
\\
\<Arrays> & \<fill\_d> (\<Object> array range)  &   5 &   1   &   5   &   1 &          
\\
\<Arrays> & \<hashCode\_a> (\<int> array)        &  10 &   0   &   1   &   1 &          
\\
\<Arrays> & \<hashCode\_b> (\<Object> array)    &  11 &   0   &   1   &   1 &          
\\
\<Arrays> & \<insertionSort\_a> (\<mergeSort>, \<Object> array) 
                                              &  15 &   3   &   1   &   2 & $\{1\}$     
\\
\<Arrays> & \<insertionSort\_b> (\<sort1>, \<int> array)
                                              &   8 &   3   &   1   &   2 & $\{1\}$     
\\
\<Arrays> & \<merge> (\<mergeSort>, \<Object> array)
                                              &  12 &   8   &   1   &   1 &             
\\
\<Arrays> & \<quicksortPartition> (\<sort1>, \<int> array)
                                              &  22 &   3   &   7   &   3 & $\{2\}$     
\\
\<Arrays> & \<vecswap>                        &  4  &   5   &   4   &   1 &             
\\
\<Collections> & \<replaceAll>                &  39 &   1   &   2   &   1 &             
\\
\<Collections> & \<reverse>                   &  15 &   1   &   1   &   1 &             
\\
\<Collections> & \<sort>                      &   9 &   2   &   3   &   1 &             
\\
\<Vector> & \<indexOf>                        &  12 &   1   &   3   &   2 &             
\\
\<Vector> & \<lastIndexOf>                    &  15 &   1   &   3   &   2 &             
\\
\<Vector> & \<removeAllElements>              &   8 &   0   &   3   &   1 &             
\\
\<Vector> & \<removeRange>                    &  12 &   1   &   3   &   1 &             
\\
\<Vector> & \<setSize>                        &  11 &   1   &   3   &   1 &             
\\
\midrule
& \textsc{total}                              & 345  &  37  &  72   &  39 &  $\{4\}$ \\
\bottomrule
\end{tabular}
\end{table*}

We evaluated the \DYNAMATE approach by running our prototype on 28 methods from the \<java.util> package of OpenJDK \JAVA 1.6.

\subsection{Case Study Selection and Preparation} \label{sec:case-study-preparation}

\autoref{table:stats-casestudy} lists the 28 methods used in the case study, which we obtained as follows.
We initially considered all methods with loops in \<java.util>: there are 421 such methods in 50 classes, for a total of 575 loops.

Since \DYNAMATE uses \ESCJ as static prover, only methods that \ESCJ can verify by \emph{manually} providing suitable loop invariants may be within \DYNAMATE's capabilities.
To identify them, we conducted a preliminary assessment where we tried to manually specify and annotate with loop invariants, and verify using \ESCJ, methods with loops of \<java.util>.
Overall, we spent about four person-weeks in the assessment process. Unsurprisingly, we spent most of the time devising the loop invariants and working around \ESCJ's limitations---precisely what \DYNAMATE can provide automation for---whereas writing pre- and postconditions was generally straightforward and quick.
Nonetheless, we had to write pre- and postcondition ourselves rather than reusing Leavens et al.'s~\cite{JDK-specs}, which include complex JML features unsupported by \ESCJ and in fact have not been used for verification of the method implementations.

During the assessment, we tried to verify every method against functional specifications as complete as possible.
In the end, we dropped all methods for which we could not get \ESCJ to work with non-trivial functional specifications within the available time budget or without resorting to complex features.
Here are some common reasons that lead us to drop methods during the assessment phases:
\begin{itemize}
\item Methods relying on language types and operations that are not adequately supported by \ESCJ's encoding and backend SMT solver: non-linear arithmetic, complex bitwise operations, floating point arithmetic, and strings.
\item Methods including calls to native code whose semantics is cumbersome or impossible to specify accurately (although we specified a few basic native methods such as \<arraycopy>).
\item Methods relying on I/O and other JVM services---most notably, reflection features.
\item Methods requiring substantial modifications to their implementation, specification, or both to be verified; in particular, we excluded methods whose verification requires ghost code or complex framing conditions.
\item Data-structure classes relying on complex class invariants whose specification is a challenge of its own~\cite{PTFM-FM14-SemiCola}.
\end{itemize}

When overloaded methods operating on different basic types were available, we included the versions operating on \<int> and on \<Object> (as placeholder for \<Integer> in sorting methods); even if the algorithms are the same, the latter typically require more complex annotations since they have to deal with \<null> values.
By the same token, we dropped variants of methods operating on different types, such as \<short> and \<byte>, when they were mere duplicates that carried no new challenges for verification.

Overall, we retained the 26 \<java.util> methods listed in \autoref{table:orig-casestudy}.
As described in the last column, we slightly modified a few of the methods to make up for features that needlessly hinder automated reasoning (e.g., external iterators and control-flow breaking instructions) or that \ESCJ does not fully support (e.g., bitwise operations and ``for each'' loops).
We also factored out the loops in \<sort1> and \<mergeSort> into separate methods according to the algorithms they implement.
Only three method modifications affected the number of loops.
Methods \<replaceAll> and \<reverse> consist of a conditional at the outermost level, with a loop in each branch that caters to the specific features of the collection such as whether it offers random access or external iterators; in our experiments, we only retained the loops (one per method) in the branch corresponding to \<Integer> indexed lists.
The single loop in \<removeLastOccurrence> visits, in reverse order, the elements of a double-ended queue implemented as a circular array.
It turns out that the asymmetry in the index range $[\mcode{head}..\mcode{tail})$ of valid elements in the queue (from index \<head> \emph{included} to index \<tail> \emph{excluded}) makes reasoning more complicated when elements are visited backward (from last to first) than when they are visited forward (from first to last, as in \<removeFirstOccurrence>).
As a result, we found no simple way to reason about \<removeLastOccurrence>'s implementation as is without the help of ghost code; in particular, reversing the approach used with \<removeFirstOccurrence> does not seem to work.
Instead, we refactored \<removeLastOccurrence>'s single loop into three different loops with the same semantics but in a form more amenable to reasoning.
None of these modifications altered the input/output semantics of the loops or the essence of the original algorithms, even though it is possible that different modifications would also work.

In all, we obtained the 28 methods listed in \autoref{table:stats-casestudy}, which we used in our case study.
The specifications we wrote also include few boilerplate frame conditions, class invariants, and definitions of exceptional behavior; since these are straightforward and do not impact the core of the correctness proofs, \autoref{table:stats-casestudy} does not list them.
We made the resulting fully annotated 28 methods obtained by this process publicly available for repeatability and for related research on verification of real Java code: 
\begin{center}
\url{https://bitbucket.org/caf/java.util.verified/}
\end{center}

\nicepar{\ESCJ as a sound verifier.}
\ESCJ does not support a number of \JAVA and \JML features~\cite{Kiniry:2006:SCW:1181195.1181200,CKC-implnotes}, such as overflow checks, inherited annotations, string literals, multiple inheritance, and static initializers, whose encoding in unsound with respect to their intended semantics.
Unsupported features are compromises that ``increase automation, improve performance,
and reduce both the number of false positives and the annotation overhead''~\cite{compromises}, and as such they are common in the design of static verifiers. 
To ensure validity of our experiments, we ascertained---by manually inspecting the source code against \ESCJ's list of documented unsupported features---that the code of our case study uses none of the unsupported \JAVA and \JML features on which the verified specifications depend.
Regarding loops, our experiments always call \ESCJ with the \verb!-loopSafe! option which, as explained in \autoref{subsec:validationofloopinvariants}, produces a sound semantics of loops thorough invariants.

\nicepar{Positioning.}
\autoref{table:stats-casestudy} gives an overview of the case study subjects, but we should keep in mind that metrics such as lines of code or specification clauses are poor indicators of the complexity of full verification of functional properties.
Even the shortest algorithms can be extremely tricky to specify and verify (see \cite{filliatre12vstte} for an extreme example), as they may require annotations involving complex disjunctions and quantifiers; the model-based approach helps express complex predicates \cite{LoopInvariantSurvey-TR-19112012} but it cannot overcome the intrinsic formidable complexity of automated reasoning.

Although the data structure implementations of our examples use arrays, the model-based style is largely oblivious of this detail, and its abstraction ensures that the general \DYNAMATE approach remains applicable in principle to other types of data structures.
Besides, even if the full formal verification of complex linked data structures such as hash tables and graphs has made substantial progress in recent years~\cite{ZeeKR08,MehnertSBS12,PTFM-FM14-SemiCola}, it remains a challenging problem for which full automation is still out of reach.
Taking stock, our case study subjects consist of real code that is representative of the state of the art of formal software verification and highlights \iflong outstanding\fi challenges to providing full automation.

\iflong
\begin{table*}[!ht]
\caption{%
\textnormal{%
Experimental results for the case study. 
Column \textsc{success rate} displays the percentage of runs that found a full correctness proof.
The numbers in the other columns are means over 30 \textsc{dynamate} runs: the percentage of \textsc{proved} proof obligations;  
the number of \textsc{iterat}ions of \textsc{dynamate};
the number of \textsc{proved} invariants (in parentheses, how many of them come from \GINDYN) and how many more would be required for a successful proof (\textsc{miss}ing); 
the number of \textsc{waves} of \GINDYN candidates considered, how many \textsc{cand}idate invariants they generated, and what percentage of the candidates were \textsc{fals}ified by dynamic analysis or removed in the \textsc{taut}ology elimination phase;
the \textsc{time} spent in \textsc{total} (seconds), and the share of each tool.%
}%
}
\label{table:results_array}
\centering
\scriptsize
\setlength{\tabcolsep}{2pt}
\begin{adjustwidth}{-1mm}{-1mm}
\begin{tabular}{rl r r r @{$\ \quad$} rr r | r r r r | r r r r r}
\toprule
& & \multicolumn{1}{c}{\textsc{success}} 
&
&
\multicolumn{1}{c}{\textsc{\#}}
&
\multicolumn{3}{c}{\textsc{\# invariants}}
&
\multicolumn{4}{|c|}{\textsc{gin-pink}}
&
\multicolumn{5}{c}{\textsc{TIME}}
\\
{\textsc{class}} & {\textsc{method}}
   & \multicolumn{1}{c}{\textsc{rate}}
   & \multicolumn{1}{c}{\textsc{proved}}
   & \multicolumn{1}{c}{\textsc{iterat.}}
   & \multicolumn{2}{c}{\textsc{proved}}
   & \multicolumn{1}{c}{\textsc{miss.}}
& \multicolumn{1}{|c}{\textsc{waves}}
   & \multicolumn{1}{c}{\textsc{cand.}}
   & \multicolumn{1}{c}{\textsc{fals.}}
   & \multicolumn{1}{c}{\textsc{taut.}}
& \multicolumn{1}{|c}{\textsc{total}}
   & \multicolumn{1}{c}{\textsc{EvoSuite}}
   & \multicolumn{1}{c}{\textsc{Daikon}}
   & \multicolumn{1}{c}{\textsc{gin-pink}}
   & \multicolumn{1}{c}{\textsc{Esc/Java2}}
\\
\midrule
\input{results_javaUtil.tex}
\bottomrule
\end{tabular}
\end{adjustwidth}
\end{table*}
\else
\begin{table*}[!ht]
\centering
\caption{%
\footnotesize
\textnormal{%
Experimental results for the case study. 
Column \textsc{success rate} displays the percentage of runs that found a full correctness proof.
The numbers in the other columns are means over 30 \DYNAMATE runs: the percentage of \textsc{proved} proof obligations;  
the number of \textsc{iterat}ions of \DYNAMATE;
the number of \textsc{proved} invariants (in parentheses, how many of them come from \GINDYN) and how many more would be required for a successful proof (\textsc{miss}ing); 
the number of \textsc{waves} of \GINDYN candidates considered, how many \textsc{cand}idate invariants they generated, and what percentage of the candidates were \textsc{fals}ified by dynamic analysis or removed in the \textsc{taut}ology elimination phase;
the \textsc{time} spent in \textsc{total} (seconds).%
}%
}
\label{table:results_array}
\centering
\footnotesize
\setlength{\tabcolsep}{2pt}
\begin{tabular}{rl r r r @{$\ \quad$} rr r | r r r r | r }
\toprule
& & \multicolumn{1}{c}{\textsc{success}} 
&
&
\multicolumn{1}{c}{\textsc{\#}}
&
\multicolumn{3}{c}{\textsc{\# invariants}}
&
\multicolumn{4}{|c|}{\textsc{\GINDYN}}
&
\multicolumn{1}{c}{\textsc{Time}}
\\
{\textsc{class}} & {\textsc{method}}
   & \multicolumn{1}{c}{\textsc{rate}}
   & \multicolumn{1}{c}{\textsc{proved}}
   & \multicolumn{1}{c}{\textsc{iterat.}}
   & \multicolumn{2}{c}{\textsc{proved}}
   & \multicolumn{1}{c}{\textsc{miss.}}
& \multicolumn{1}{|c}{\textsc{waves}}
   & \multicolumn{1}{c}{\textsc{cand.}}
   & \multicolumn{1}{c}{\textsc{fals.}}
   & \multicolumn{1}{c}{\textsc{taut.}}
& \multicolumn{1}{|c}{\textsc{secs.}}
\\
\midrule
\input{results_javaUtil_short.tex}

\bottomrule
\end{tabular}
\end{table*}
\fi

\nicepar{Experimental setup.}
All experiments were performed on a Ubuntu GNU/Linux system installed on a 2.0 GHz Intel Xeon processor and 2 GB of RAM.
We set a timeout of 120 seconds to every invocation of \EVOSUITE and of 180 seconds to every invocation of \ESCJ.
As \EVOSUITE uses randomized algorithms, its results are not deterministic; to control for this, we repeated every complete experiment 30~times and report the mean of every measure.
\DYNAMATE's ran on the 28 methods of \autoref{table:stats-casestudy} annotated with pre- and postconditions but without any loop invariants: the goal of the experiments was for \DYNAMATE to infer loop invariants without additional input.

\subsection{Experimental Results} \label{sec:evaluation-results}

\autoref{table:results_array} summarizes the experimental results, where
\DYNAMATE automatically verified 25 of the 28 methods in at least one of the 30 repeated runs (indicated by a success rate above 0\%). 

A success rate below 100\% indicates methods for which \EVOSUITE performed below par, and thus \DYNAMATE reached a fix point in invariant discovery in some of the 30 repeated runs.
Specifically, \EVOSUITE may fail to find counterexamples to invalid invariants, which thus remain in the set $I \setminus \|INV|$ of candidates neither conclusively falsified nor conclusively proved.
This may prevent more general invariants from being inferred, thus jeopardizing the whole verification effort; for example, a surviving invalid invariant $\mcode{i} \in \{\mcode{0}, \mcode{1}, \mcode{2}\}$ would shadow a valid invariant $\mcode{i} \geq \mcode{0}$.
Nonetheless, experiments with \DYNAMATE normally have high repeatability: 21 of the 25 verified subjects have a success rate above 50\%.

\begin{result}
\DYNAMATE automatically verified 25 of the 28 subjects, with high repeatability. 
\end{result}

\begin{figure}
\begin{lstlisting}
merge:
   destHigh - ic == (mid - p) + (high - q)

quicksortPartition:
   b <= c ==> x[b] > v

sort:
   \forall int j: 0 <= j && j < a.length ==> a[j] == \old(a[j])
   \forall int j: 0 <= j && j < ic ==> 
                   a[j] == l.getElementData()[j]
\end{lstlisting}
\caption{The four necessary loop invariants that \DYNAMATE failed to infer in all runs.}
\label{fig:non-inferred-invs}
\end{figure}

The success rate was 0\% for methods \<Arrays.merge>, \<Arrays.quicksortPartition>, and \<Collections.sort>.
For these methods, \DYNAMATE failed to infer all necessary invariants.
The four missing invariants, one for each of \<merge> and \<quicksortPartition> and two for \<sort>, are shown in \autoref{fig:non-inferred-invs}; they have a form that is neither among \DAIKON's templates nor \GINDYN's mutants:
\begin{itemize}
\item \<merge>'s missing invariant is an equality relation that is not among \DAIKON's invariant because it involves six variables but, by default, \DAIKON restricts templates to three variables for scalability.
\item \<quicksortPartition>'s missing invariant is also a scalar relation, but \DAIKON's templates do not instantiate disjunctive (including implication) templates without additional user input (splitting predicates).
\item \<sort>'s missing invariants are quantified expressions of the kinds that \GINDYN supports. However, \<sort>'s postcondition has a form that is unrelated to the two invariants, due to \<sort>'s peculiar implementation: a call to a native method does the actual sorting of array \<a>, and is followed by a loop that copies the result from \<a> into \<l>. Thus, the loop's invariants are unrelated to sorting, which is just a property carried over by copying.
\end{itemize}
We repeated the experiments by manually adding the missing invariants; as expected, \DYNAMATE successfully verified the methods.

Besides full functional verification, we also measured the percentage of proof obligations \DYNAMATE was able to discharge.
We counted  367 proof obligations constructed by \ESCJ for the annotated \JAVA programs used in our evaluation; 
these include all pre- and postcondition checks, the class invariant checks, and also implicit checks for out-of-bound array accesses and \<null> dereferencing.
\DYNAMATE managed to find a set of loop invariants such that 97\% of the proof obligations were discharged on average.
This dramatically decreases the number of loop invariants users have to write manually.

\begin{result}
On average, \DYNAMATE automatically discharged \\ 97\% of the proof obligations.
\end{result}

Using exclusively \DAIKON's invariants, \DYNAMATE could verify only \<hashCode\_a> and \<hashCode\_b>; this shows the importance of  \GINDYN to add flexibility to \DYNAMATE.
The other columns about \GINDYN in \autoref{table:results_array} show the crucial role of dynamic analysis and tautology elimination to discard invalid and irrelevant mutants: while postcondition mutation generated nearly 2500 mutants on average (way too many for \ESCJ to handle), dynamic analysis and tautology elimination cut this number down to below 300 (i.e., $2410 \times (1 - 0.82) \times (1 - 0.34)$).

On average, 66\% of \DYNAMATE's execution time is spent running \EVOSUITE, 15\% in \GINDYN, 14\% in \ESCJ and 6\% in \DAIKON. 
\DYNAMATE's average running time per method (45 minutes) is high compared to other dynamic techniques.
There are ample margins to optimize the \DYNAMATE prototype for better speed; in particular, using third-party components as black boxes is an obvious source of inefficiency%
\iflong
(\EVOSUITE, for example, would be much faster if we customized the fitness function or seeded initialization)\fi%
.
At the same time, \DYNAMATE solves an intrinsically complex task---fully automated correctness proof without loop an\-no\-ta\-tions---compared to the standard goal of dynamic techniques; given the state of the art, we should not have unrealistic expectations regarding its performance.

\subsection{Experimental Comparison} \label{sec:comparison}

The staggering amount of research on loop invariant generation and automated verification has produced a wide variety of results that are not always directly comparable.
In this section, we report on focused experiments involving few cutting-edge tools that can be directly compared to \DYNAMATE, to convincingly show how it improves the state of the art and its specific strengths and weaknesses.

\nicepar{Tool selection.}
We selected other tools for automated verification that are in the same ``league'' as \DYNAMATE; namely, they satisfy the following characteristics:
\begin{inparaenum}[\itshape i)]
\item
a working implementation is available;
\item
they work on a real programming language, or at least a significant subset;
\item
they support numerical and functional properties;
\item
they are completely automatic (or we clarify what extra input is needed); 
\item
they are cutting-edge (i.e., no other similar tool supersedes them).
\end{inparaenum}
The first two requirements excluded several techniques without serviceable implementations, or that only work on toy or highly specialized examples; we provide more details about some of the most significant exceptions at the end of the section.
All in all, we identified three ``champions'' with these characteristics: \INVGEN~\cite{gupta2009+}, which uses constraint-based techniques and mainly targets linear numeric invariants; \BLAST~\cite{blast}, a software model-checker using CEGAR and  predicate abstraction; and \CCC~\cite{DBLP:conf/foveoos/FahndrichL10}, the CodeContracts static checker formerly known as Clousot~\cite{CousotCL11}, based on abstract interpretation with domains for arrays.

We translated the classes used in \DYNAMATE's evaluation (including the specification predicates in \<TArrays> and \<TLists>) into a form amenable to each tool: to C for \INVGEN and \BLAST, and to C\# for \CCC.
We tried to replicate the salient features of \JAVA's semantics in each language, without introducing unnecessary complications; for example, \JAVA\ \<Object>s become C \lstinline[language=C]|struct|s passed to functions as pointer arguments.

We encoded in each tool as many of the \ESCJ proof obligations as possible.
We used \<assert>s to express proof obligations for which no equivalent language feature was available (for example, class invariants in \BLAST). 
When a tool's specification language did not support quantified expressions (as is the case in C Boolean expressions), we rendered the semantics using operational formulations, like \DYNAMATE does for runtime checking (\autoref{sec:evosuite}).

For each tool, we measured the percentage of the 367 proof obligations that was expressible and the percentage that was successfully verified. 
We also include figures about the bare \ESCJ, which can discharge a good fraction of the proof obligations that do not require reasoning about loops. 
\autoref{tab:comparison-results} shows the results.

\begin{table}
\begin{center}
\scriptsize
\begin{tabular}{l r r r r r }
\toprule
\multicolumn{1}{c}{\textsc{tool}} & \multicolumn{2}{c}{\textsc{proof obligations}} & \multicolumn{1}{c}{\textsc{verified}}  & \multicolumn{1}{c}{\textsc{time}}   \\
\multicolumn{1}{c}{\textsc{}} & \multicolumn{1}{c}{\textsc{expressible}} & \multicolumn{1}{c}{\textsc{proved}}  & \multicolumn{1}{c}{\textsc{methods}} &  \\
\midrule
\ESCJ     & $100\,${\scriptsize \%}   &  231 ($63\,${\scriptsize \%}) & 0/28 &$122\,${\scriptsize s} \\
\hline
\INVGEN   & $42\,${\scriptsize \%}    &  60  ($39\,${\scriptsize \%}) &  0/28 & $78\,${\scriptsize s}    \\
\BLAST    & $100\,${\scriptsize \%}   &  238 ($65\,${\scriptsize \%}) & 3/28  & $5431\,${\scriptsize s}  \\
\CCC      & $100\,${\scriptsize \%}   &  276 ($75\,${\scriptsize \%}) & 3/28 & $106\,${\scriptsize s}   \\
DYNAMATE  & $100\,${\scriptsize \%}   &  354 ($97\,${\scriptsize \%}) & 25/28 & $75348\,${\scriptsize s} \\
\bottomrule
\end{tabular}
\end{center}
\caption{Experimental comparison of \DYNAMATE against other tools for automated verification.}
\label{tab:comparison-results}
\end{table}

\nicepar{Comparison results.}
\ESCJ gives a good baseline of 63\% proof obligations automatically discharged.
\INVGEN is quite limited in what it can express: the subset of C it inputs does not have support for arrays, and hence it is strictly limited to scalar properties; within these limitations, it is often successful using its predefined templates.
\BLAST supports the full C language, but does not go much beyond the baseline in terms of what it can prove. In fact, ``the predicates tracked by \BLAST do not contain logical quantifiers. Thus, the language of invariants is weak, and \BLAST is not able to reason precisely about programs with arrays or inductive data structures whose correctness involves quantified invariants''~\cite{blast}.
\CCC's performance is impressive, also given its full support of .NET, but its abstract domains are still insufficient to express complex quantifications over arrays formalizing, for example, sortedness.
\DYNAMATE achieves a solid 97\% of automatically discharged proof obligations, significantly improving over the state of the art: the proof obligations discharged by \DYNAMATE are a superset of those checked by other tools.\footnote{With the sole exception of a property in \<Collections.sort>, which \BLAST can check using its context-free option \!-cfb!. However, \BLAST can also check the negation of the same property, which indicates unsoundness; hence, this exception seems immaterial for the comparison.}
In particular, \DYNAMATE achieved full verification of $25$ out of $28$ methods, while the other tools verified at most $3$ methods. 

\begin{result}
\DYNAMATE automatically verified 28\% more proof obligations than state-of-the-art verification tools.
\end{result}

Given that our selection of 28 methods is based on their analyzability with \ESCJ (as we describe at the beginning of \autoref{sec:evaluation}), we do not claim that our results generalize to other kinds of programs or properties such as pointer reachability. They do, however, provide evidence that dynamic techniques can complement static ones in order to automate full verification of ``natural'' programs that are currently challenging for state-of-the-art tools. (In particular, our \DYNAMATE implementation boosts the capabilities of \ESCJ.)

The main weakness of \DYNAMATE derives from its usage of dynamic randomized techniques: \DYNAMATE may require repeated runs, and is one to two orders of magnitude slower than the static tools\iflong, which normally terminate analysis in a matter of minutes\fi.
Future work will target these shortcomings to further promote the significant results obtained by \DYNAMATE's algorithms.
Also part of future work is evaluating \DYNAMATE on examples originally used to evaluate \INVGEN, \BLAST, or \CCC; and integrating in \DYNAMATE other third-party tools tailored to \emph{some kinds} of invariants.

\nicepar{Other tools.}  The following table summarizes the crucial
features that distinguish \DYNAMATE from a few other cutting-edge tools.
Only JPF~\cite{pasareanu2004} is available (\textsc{a/u}) but is not directly comparable as
it is limited to bounded checking (no exhaustive verification). Based on
\cite{HoderKV11,SrivastavaG09}, the other techniques also have
limitations; see \autoref{sec:related-work} for qualitative details.

\begin{center}
\scriptsize
\begin{tabular}{l c l}
\toprule
\multicolumn{1}{c}{\textsc{tool}} & \multicolumn{1}{c}{\textsc{a/u}} & \multicolumn{1}{c}{\textsc{limitations}} \\
\midrule
\JAVA Pathfinder~\cite{pasareanu2004}     &  A  &  bounded symbolic execution \\
Vampire~\cite{HoderKV11}             &  U  &  linear array access, no nesting \\
Srivastava et. al~\cite{SrivastavaG09} &  U  &  requires templates and predicates \\
\bottomrule
\end{tabular}
\end{center}

\subsection{Threats to Validity}
\label{sec:threats}

\emph{External validity} concerns the generalization of our results to subjects other than the ones we studied.  
Regarding specifications, although we followed a particular style (the model-based approach), the style is general enough that it remains applicable to other software~\cite{PFPWM-ICSE13}.
\iflong
At this point, we do not expect \DYNAMATE to scale to programs that are significantly larger or much more complex than the ones we have analyzed. 
\fi
In addition, we wrote specifications as complete as possible with respect to the full functional behavior that can be manually proved with reasonable effort using \ESCJ; this makes our case study a relevant representative of the state of the art in formal software verification.
Targeting \DYNAMATE's approach to the verification of properties other than functional, as well as tackling significantly larger programs, belongs to future work.

We took all measures necessary to minimize threats to \emph{internal validity}---which originate in the execution of experiments.
To minimize the effect of chance due to the usage of randomized algorithms (\EVOSUITE), we repeated each experiment multiple times and considered average behavior; in most cases, chance negatively affected only a small fraction of the runs. 
We also manually inspected all results (inferred loop invariants), which gives us additional confidence about their internal consistency.
\DYNAMATE and its modules depend on parameters such as weights, timeouts, and thresholds which might influence experimental results. 
We ran all subjects using the same parameters, for which we picked default values whenever possible.

Threats to \emph{construct validity} have to do with how appropriate the measures we took are.
We measured success in terms of how many proof obligations \DYNAMATE discharged automatically. 
This certainly is a useful metric; assessing other measures such as execution time required per proof obligation belongs to future work.

\section{Related Work} \label{sec:related-work}

Since we cannot exhaustively summarize the huge amount of
work on automating program verification indirectly related to \DYNAMATE,
we focus this section on the problem of inferring loop invariants to
automate functional verification; a natural classification is in
static and dynamic techniques.  
\autoref{sec:comparison} presents a
more direct comparison between \DYNAMATE and a few selected
``champions'' of loop invariant inference.

\subsection{Static Techniques}

\nicepar{Abstract interpretation}~\cite{CC77} is a general framework
for computing sound approximations of program semantics; invariant
inference is one of its main applications.  Each specific abstract
interpreter works on one or more \emph{abstract domains}, which
characterize the kinds of invariants that can be computed.  Much of
the work in abstract interpretation has focused on domains for
non-functional properties, such as pointer reachability~\cite{NikolicS13} and
other heap shape properties~\cite{CL05}\, or on simple ``global''
correctness properties~\cite{BCC+03-Astree,wala}, such as absence of
division by zero or null pointer dereference. 
Domains exist for simple numerical properties such as polynomial equalities and
inequalities~\cite{ppl,BagnaraHZ08SCP} and interval (bounding)
constraints involving scalar variables~\cite{CH78,apron,JeannetM09}; these
do not include quantified array expressions, which are needed to
express the functional correctness of several programs used in
\DYNAMATE's evaluation, such as searching and sorting algorithms. 
Only recently have the first abstract domains supporting restricted
quantification over arrays been developed~\cite{gulwani2008,CousotCL11}. 

Compared to \DYNAMATE, these techniques offer much more scalability and speed, but also incur some
limitations in terms of language support and flexibility. 
In particular, Gulwani et al.~\cite{gulwani2008} require user input in the form of
templates that detail the structure of the invariants, and its
experiments do not target nested loops (e.g., only inner loops of
searching algorithms). The technique behind the CodeContracts static
checker~\cite{CousotCL11} also cannot deal with some quantified
properties that \DYNAMATE can handle (see \autoref{sec:comparison} for a direct comparison to  \CCC~\cite{CousotCL11}). More generally, abstract
interpretation is limited to domains fixed a priori and may lose
precision in the presence of unsupported language features, whereas
\DYNAMATE (largely thanks to its reliance on dynamic analysis) works
in principle with any property expressible through \JML annotations and
arbitrary Java implementations.

\nicepar{Predicate abstraction}~\cite{graf1997,flanagan2002} is a technique to build
finite-state over-approximations of programs, which can be seen as a
form of abstract interpretation. The applicability of predicate
abstraction crucially depends on the set of predicates provided as
input, which determine precision and complexity.  

Whereas predicates may be collected from the program text following some heuristics 
\cite{jung2010,schmitt2007} or they may be manually specified through
annotations \cite{weiss2010,flanagan2002}, \DYNAMATE requires no input
besides a specified program.  
However, the two techniques 
may be usefully combined, with the invariants guessed by \DYNAMATE used to
build a predicate abstraction.

Predicate abstraction is a fundamental component of the CEGAR
(Counter-Example Guided Abstraction Refinement)
approach~\cite{cegar-clarke} to software model-checking, which
features in tools such as SLAM~\cite{slam} and \BLAST~\cite{blast}.
Software model checkers specialize in establishing state reachability
properties or other temporal-logic properties, whereas they hardly
handle complex invariants involving quantification for functional
correctness (such as those central to \DYNAMATE's evaluation), as we
demonstrate in \autoref{sec:comparison}.

\nicepar{Constraint-based techniques} reduce the invariant inference
problem to solving constraints over a template that defines the
general form of invariants (playing a somewhat similar role to
abstract domains in abstract interpretation). 
The challenge of developing new constraint-based techniques lies in defining
expressible yet decidable logic fragments, which characterize
extensive template properties.  

The state of the art focuses on
invariants in the form of Boolean combinations of linear~\cite{colon2003,gupta2009+,GulwaniSV09}
and quadratic~\cite{GulwaniSV08} inequalities,
polynomials~\cite{sankaranarayanan2004,Kapur05,RCK07},
 restricted properties of arrays~\cite{BHIKV09} and matrices~\cite{HHKV10}, 
and linear arithmetic with uninterpreted functions~\cite{beyer2007}.
Recent advances include automatically computing least fixed points of
recursively defined predicates
\cite{DBLP:conf/sat/HoderB12,DBLP:conf/vmcai/Bradley11}, which are can
express the verification conditions of transition systems modeling
concurrent behavior.

Since constraint-based techniques rely on decidable logic fragments,
 they rarely support templates involving quantification.  The
automata-based approach of~\cite{BHIKV09} is an exception, but its
experiments are still limited to flat linear loops with simple control
logic, and it is not applicable to linked structures.

See \autoref{sec:comparison} for a direct comparison to \INVGEN~\cite{gupta2009+}.

\nicepar{Using first-order theorem proving.}  
Kov{\'a}cs et al.~\cite{kovacs2009,HoderKV11} target the inference of
loop invariants for array-ma\-nip\-u\-lat\-ing programs.  Their technique
encodes the semantics of loops directly as recurrence relations and
then uses a properly modified saturation theorem prover to derive
logic consequences of the relations that are syntactically loop
invariants.  This can generate invariants involving alternating
quantifiers, which are out of the scope of most other techniques.
McMillan~\cite{mcmillan2008} also uses a modified saturation theorem
prover to infer quantified loop invariants describing arrays and
linked lists.  

Both techniques rely on heuristics that substantially depend on
interacting with a custom-modified theorem prover, which limits their
practical applicability to programs with ``regular'' behavior.  For
example, \cite{kovacs2009} assumes loops that monotonically increment
or decrement a counter variable; more general index arithmetic is not
handled.  As a result, the example programs demonstrated in
\cite{kovacs2009,HoderKV11,mcmillan2008} are limited to linear access
to array elements and no nested loops, and do not include anything as
complicated as sorting (also see \autoref{sec:comparison}).  
While more complex quantified invariants could be generated in principle, doing so normally requires\footnote{Laura Kov\'acs, personal communication, 14 March 2014.} massaging the input programs into a form amenable to the ``regularity'' assumptions on which the inference technique relies.
In contrast, \DYNAMATE uses a static prover and other components as
black-boxes, and works on real implementation of sorting and searching
algorithms.  

Totla and Wies~\cite{TotlaW13} apply interpolation
techniques to axiomatic theories of arrays and linked lists, which
supports invariant inference over expressive domains; however, the
examples the technique can handle in practice are still limited to flat loops and
straightforward linked list manipulations (the technique for arrays has not
been implemented).

\nicepar{Combination of static techniques.}
\HAVOC~\cite{lahiri2009} pioneered using a static verifier to check if candidate assertions are valid: 
it creates an initial set of candidates (possibly including loop invariants) by applying a fixed set of rules to the available module-level contracts (i.e., module invariants and interface specifications).
Like \DYNAMATE, \HAVOC applies the \HOUDINI algorithm to determine which candidates are valid.
Using only static techniques, however, may introduce false negatives, that is valid loop invariants being erroneously discarded.
In contrast, \DYNAMATE discards a large number of invalid candidates by runtime checking, which is immune to false negatives.
Static verification is only applied later and, if the program verifier fails, \DYNAMATE will enter another iteration of test case generation, trying to bring in additional precision.

JPF (\JAVA Pathfinder) supports heuristics~\cite{pasareanu2004} to iteratively strengthen loop invariants from path conditions and intermediate assertions.\footnote{The technique of \cite{pasareanu2004} was not fully implemented in JPF at the time of writing (Willem Visser, personal communication, 18~February~2014).}  
While JPF mainly targets concurrency errors such as race conditions, these heuristics are also applicable to numeric invariants.
However, as the \DAIKON experience shows, invariants may not be apparent from the code (e.g., $\mcode{x} < \mcode{y}$ might not appear in the code but be a necessary loop invariant).
JPF can also leverage postconditions if they are available, but without mutating them as \DYNAMATE does; furthermore, 
JPF's symbolic-execution approach cannot handle the kind of complex verification conditions that program provers such as \ESCJ fully support and, more crucially, is limited to bounded state spaces (such as arrays of bounded size).

\HOLA~\cite{DilligDLM13} implements an inference technique somewhat similar to JPF, but uses abduction (a mechanism to infer premises from a given conclusion) for strengthening, which gives it more flexibility and generality. 
However, it 
is also limited to invariants consisting of Boolean combinations of linear integer constraints  (inequalities and modular relations).

Srivastava and Gulwani~\cite{SrivastavaG09} combine predicate abstraction and template-based inference to construct quantified invariants that can express complex properties such as sortedness.
Their tool takes as input a program and a set of templates and predicates; 
for example, inferring invariants specifying sortedness of an array \<A> requires a template $\forall k: \Box \Longrightarrow \Box$ and predicates $\mcode{0} \leq \mcode{x}$, $\mcode{x} < \mcode{y - 1}$ and $\mcode{A[}k\mcode{]} \leq \mcode{A[}k \:\mcode{+ 1]}$.
They demonstrate the approach on several sorting algorithms (as well as simple linked list operations)\iflong, where it computes the invariants to prove full functional correctness in few seconds\fi.
While these results are impressive, \DYNAMATE still offers some complementary advantages (also see \autoref{sec:comparison}): it works on real implementations and supports the full Java programming language; by using model-based annotations, it is more flexible with respect to the used data structures (e.g., lists vs.\ arrays); and does not require extra user input in the form of templates and predicates.
In fact, the \DYNAMATE approach could also accommodate templates to suggest invariant shapes, thus dramatically narrowing down the search space; conversely, Srivastava and Gulwani's technique could be extended to generate templates from postconditions as \DYNAMATE's \GINDYN does.

\nicepar{Separation logic}
is an extension of Hoare logic specifically designed to specify and reason about linked structures in the heap~\cite{Reynolds02}. Consequently, the bulk of the work in invariant inference based on separation logic focuses on pointer reachability and other shape properties~\cite{Mag+06,MagillTLT10,VogelsJPS11,CalcagnoDOY11}, often by means of abstract interpretation techniques using domains specialized for such properties.
Program verifiers based on separation logic also target similar properties, and typically provide a level of automation that is somewhat intermediate between that of ``push-button'' tools (such as \DYNAMATE) and of interactive provers (such as Isabelle and Coq).
For example, \JSTAR is a powerful separation-logic verifier that works on real Java code and includes support for loop invariant inference~\cite{DistefanoP08}; it focuses on heap reachability properties\iflong\footnote{Indicatively, the \JSTAR tutorial (\url{http://www.jstarverifier.org/jstar.tutorial.pdf}) analyzes linked lists but does not prove any functional properties of their methods.}\fi, requires auxiliary annotations in the form of so-called abstract predicates and custom inference rules, and falls back to user interaction when automated inference is not successful.
For these reasons, we did not include separation-logic based tools in our comparison with \DYNAMATE.
In fact, extending \DYNAMATE's techniques to work with separation-logic properties is an interesting direction for future work.

\nicepar{}
Fully automatic verification requires attacking from multiple angles.  
In this respect, the \DYNAMATE architecture is flexible and can accommodate and benefit from other static approaches.

\subsection{Dynamic Techniques}

The \GUESSANDCHECK \cite{sharma2012} algorithm infers invariants in the form of algebraic equalities (polynomials up to a given degree).
\GUESSANDCHECK achieves soundness and completeness by targeting a very restricted programming language where all expressions are of Boolean or real type.
It starts from concrete program executions, which determine constraints solved using a linear algebra solver; the solution may establish a valid loop invariant or determine a counterexample that can be used to construct new executions. 
While the overall structure of \GUESSANDCHECK has some similarities to ours, \DYNAMATE targets general-purpose programs, which requires very different techniques.

The work on \DAIKON~\cite{ErnstCGN2001:TSE} pioneered using dynamic techniques to infer invariants, and has originated a lot of follow-up work. 
The bulk of it targets, however, assertions such as pre- and postconditions and not loop invariants.
Nguyen et al.'s dynamic inference technique~\cite{nguyen2012} is a noticeable exception, which generates loop invariants in the form of polynomial equations over program variables.

A general limitation of dynamic invariant inference is its reliance on predefined templates, which restricts the kinds of invariants that can be inferred.
\DYNAMATE uses the \GINDYN approach to work around this limitation: a method's postcondition suggests the possible forms loop invariants may take.
The \DYNAMATE architecture could also integrate dynamic invariant inference tools with richer or more specialized templates than \DAIKON.

\subsection{Hybrid Techniques}

Recent work in the context of CEGAR techniques has combined static verification and test case generation.
The \SYNERGY algorithm \cite{DBLP:conf/sigsoft/GulavaniHKNR06} avoids unnecessary abstraction refinements guided by concrete inputs generated using directed automatic random testing~\cite{DBLP:conf/pldi/GodefroidKS05}. 
The \DASH algorithm \cite{DBLP:conf/issta/BeckmanNRS08} builds on \SYNERGY to handle programs with pointers without whole-program may-alias analysis.
Unlike \DYNAMATE, these techniques aim at type-state properties (e.g., correct lock usage or absence of resource leaks).
Yorsh et al.~\cite{DBLP:conf/issta/YorshBS06} follow a similar technique of refining abstractions based on the behavior in concrete executions.
A model generator determines new concrete states whose execution leads to an unexplored abstract state.
The new concrete states could be unreachable, since they are derived based on the abstraction; therefore, they are not as precise as actual tests.

With the overall goal of improving symbolic execution, Godefroid and Luchaup~\cite{godefroid2011} suggest to guess loop invariants from concrete execution traces; the invariants are then used to summarize loop executions when constructing path conditions.  
The \DYSY approach~\cite{csallner2008} directly mines invariants from the collected path conditions with no required predefined invariant patterns; like \DAIKON, it produces a collection of likely invariants, which may include unsound ones; \DYSY's constraint-based techniques, however, may provide more flexibility in terms of the invariant forms that can be mined.

Nimmer and Ernst~\cite{NimmerE01:RV,nimmer2002} combine dynamic invariant inference \`a la Daikon with a static program checker.
Loop invariants are out of the scope of that work, since they use ESC/Java with unsound loop approximation (i.e., single unrolling); hence, ``loop invariants may need to be strengthened to be proved''~\cite{NimmerE01:RV}. 
Nguyen et al.~\cite{nguyen2014} perform static inductive validation of dynamically inferred program invariants. 
Their work targets scalar numerical disjunctive invariants (precisely, expressible as inequalities between maxima of scalar terms), which are useful for numerical programs but cannot express some more complex functional properties such as sortedness.
\DYNAMATE not only provides loop invariants for full program proofs; it also closes the cycle by means of a test input generator, which makes the overall technique completely automatic.

Nori and Sharma~\cite{Nori013} use a program verifier in combination with automatic test case generation to automatically produce termination proofs.
The kinds of loop invariants required for termination proofs are simpler than those discovered by \DYNAMATE, since they only have to constrain the values of a ranking function---an integer expression showing progress.
Thus, invariants based on predefined templates are sufficient for termination but are only a small ingredient of \DYNAMATE.

\section{Conclusions and Future Work}
\label{sec:conclusion}

We have presented a fully automated approach to discharge proof obligations of programs with loops.
The approach combines complementary techniques: test case generation, dynamic invariant detection, and static verification.
A novel and important component of our work is a new fully automatic technique for loop invariant detection based on syntactically mutating postconditions.   
Our \DYNAMATE prototype automatically discharged 97\%~of all proof obligations for 28 methods with loops from \<java.util> classes. 

Besides general improvements such as scalability and performance, our future work will focus on the following issues:

\textbf{Better test generators:}  As any module in \DYNAMATE can be replaced by a better implementation of the same functionalities, we are currently investigating dynamic/symbolic approaches to test case generation~\cite{DBLP:conf/tap/JamrozikFTH13} as well as hybrid techniques integrating search-based and symbolic approaches~\cite{malburg+fraser-ase}.

\textbf{More diverse invariant generators:}   We are exploring \emph{evolutionary} approaches in which invariants are systematically evolved from a grammar over a small set of primitives~\cite{ratcliff2011}.
We also plan to apply techniques based on symbolic execution (such as the one implemented in \DYSY~\cite{csallner2008}) to provide for more, and more diversified, loop invariant candidates.

\textbf{Stronger component integration:}  If a proof fails, a program verifier may be able to produce a \emph{counterexample}, which would make an ideal input to the test case generation module for a new iteration. We are researching how to leverage such additional information, whenever available, while preserving the low coupling of \DYNAMATE's architecture.

\DYNAMATE can become a platform on which several approaches to test generation, dynamic analysis, and static verification can work in synergy to produce a greater whole.  
We are committed to make the \DYNAMATE framework publicly available, including all subjects required to replicate the results in this paper.  For details, see:

\begin{center}
\url{http://www.st.cs.uni-saarland.de/dynamate/}
\end{center}

\section*{Acknowledgments}
The research leading to these results has received funding from the European Research Council under the European Union's Seventh Framework Programme (FP7/2007-2013) / ERC grant agreement nr.~[290914] and EU FP7 grant 295261 (MEALS).
The second author was partially funded by the Swiss SNF Grant ASII 200021-134976.
Klaas Boesche, Alessandra Gorla, Jeremias R\"o\ss{}ler, and Christoph Weidenbach provided helpful comments on earlier revisions of this work.

\bibliographystyle{IEEEtran}
\bibliography{IEEEabrv,dynamate-journal}


\end{document}

%% file: jml-listings.tex
\lstdefinelanguage[JML]{Java}[]{Java}%
       {
        comment=[l]{//\ },
        morecomment=[s]{/*\ }{*/},        
        morecomment=[s]{/**}{*/},
        classoffset=1,
        morekeywords={abrupt_behavior,abrupt_behaviour,
         accessible,accessible_redundantly,also,assert,assert_redundantly,
         assignable,assignable_redundantly,assume,assume_redundantly,
         axiom,behavior,behaviour,breaks,breaks_redundantly,
         callable,callable_redundantly,captures,captures_redundantly,
         choose,choose_if,code,code_bigint_math,code_java_math,
         code_safe_math,constraint,constraint_redundantly,constructor,
         continues,continues_redundantly,decreases,decreases_redundantly,
         decreasing,decreasing_redundantly,diverges,diverges_redundantly,
         duration,duration_redundantly,ensures,ensures_redundantly,
         example,exceptional_behavior,exceptional_behaviour,
         exceptional_example,exsures,exsures_redundantly,extract,field,
         forall,for_example,ghost,helper,hence_by,hence_by_redundantly,
         implies_that,in,in_redundantly,initializer,initially,instance,
         invariant,invariant_redundantly,loop_invariant,
         loop_invariant_redundantly,maintaining,maintaining_redundantly,
         maps,maps_redundantly,measured_by,measured_by_redundantly,method,
         model,model_program,modifiable,modifiable_redundantly,modifies,
         modifies_redundantly,monitored,monitors_for,non_null,
         normal_behavior,normal_behaviour,normal_example,nowarn,
         nullable,nullable_by_default,old,or,post,post_redundantly,
         pre,pre_redundantly,pure,readable,refine,refines,refining,represents,
         represents_redundantly,requires,requires_redundantly,
         returns,returns_redundantly,set,signals,signals_only,
         signals_only_redundantly,signals_redundantly,spec_bigint_math,
         spec_java_math,spec_protected,spec_public,spec_safe_math,
         static_initializer,uninitialized,unreachable,weakly,
         when,when_redundantly,working_space,working_space_redundantly,
         writable
        },
        morekeywords={rep,peer,readonly},
        keywordsprefix=\\,
        otherkeywords={<:,<-,->,..,<==,==>,<==>,<=!=>},
        classoffset=0, 
        escapechar=\#,
          literate=%
          {:}{$\colon$}1
          {!}{$\lnot$}1
          {-}{$-$}1
          {+}{$+$}1
          {==}{$=$}1
          {!=}{$\neq$}1
          {&&}{$\land$}1
          {||}{$\lor$}1
          {<}{$<$}1
          {<=}{$\le$}1
          {>}{$>$}1
          {>=}{$\ge$}1
          {==>}{$\Longrightarrow$}3
          {<==}{$\Longleftarrow$}3
          {<==>}{$\Longleftrightarrow$}4
          {\\forall}{$\forall$}1
          {\\exists}{$\exists$}1
          {lambda}{$\lambda$}1
}

%% file: waves.tex
The current implementation defines 16 waves;  \GINDYN  executes one or more waves in each iteration of \DYNAMATE's main loop (\autoref{fig:dynamate}).
The waves are of three kinds: 
the first kind includes 7 waves 
that all work by substituting integer expressions in any postcondition \emph{clause};
the second kind includes 4 waves 
that all work by substituting integer expressions in any \emph{predicate}, in negated or unnegated form, appearing in the \emph{postcondition};
the third kind includes 5 waves 
that all work by substituting integer expressions in any predicate, in negated or unnegated form, belonging to the same \emph{collection} (\<TArrays> or \<TLists>) as those in the postcondition.
Waves of the same kind differ in how many integer expressions each generated mutant may contain (from one up to three), in whether only parameterless integer expressions are available for substitution (variables and parameterless method calls), and in whether aging or weakening are applied.
\autoref{fig:waves-example} describes one wave of each kind: the first and the second wave produce the mutants in \autoref{fig:mutations-useless} and \autoref{fig:mutations-useful}.

\begin{figure}[!htb]
\begin{center}
\footnotesize
\begin{tabular}{rc}
\textsc{kind} & \textsc{wave} \\
\midrule
1st  & $\muts \gets Q$ ; \!int! sub.\ ; \!int! sub.  \\
2nd  & $\muts \gets \{ p \in Q \cap \mcode{boolean} \}$ ; \!int! sub.\ ; aging \\
3rd  & $\muts \gets \{ \mcode{T.p} \in \mcode{boolean} \mid \exists q\colon \mcode{T.}q \in Q \}$ ; \!int! sub.\ ; aging 
\end{tabular}
\end{center}
\caption{One example mutation wave of each kind.}
\label{fig:waves-example}
\end{figure}

The three kinds of waves try to cover the most common patterns found in algorithms~\cite{LoopInvariantSurvey-TR-19112012}: mutations of the postcondition (1st kind), of some predicate in the postcondition (2nd kind), of some predicate of the same family of those in the postcondition (3rd kind).

Then, different waves of the same kind achieve different trade-offs between combinatorial complexity and exhaustiveness of the applied substitutions.
If we disregarded generation time completely, we would retain only the most general wave of each kind (for example, including as many three substitutions per mutant as well as aging and weakening) and run it to completion.
Unfortunately, this would require an exorbitant amount of time in most cases and, even in the cases where it would not lead to combinatorial explosion, it would generate too many irrelevant mutants.
Instead, we introduce waves that are incrementally more complex, so that the more amenable algorithms do not require to run the most complex waves at all.
With the current implementation of \GINDYN, this solution has the drawback of introducing redundancy in the generation of mutants, where some later waves generate mutants that were already generated by previous waves (and hence are immediately discarded).
Implementing a mutation generation process that avoids redundancies belongs to future work (although the results in \autoref{sec:evaluation} indicate that mutation generation is not the bottleneck in \DYNAMATE).

\DYNAMATE's waves target substitutions only\footnote{Note that substitutions of Boolean expressions are subsumed by the process of predicate extraction.} of integer (\<int> and \<Integer>) expressions because they are at the heart of so many algorithms and their functional specifications~\cite{LoopInvariantSurvey-TR-19112012}
Applying \DYNAMATE to disparate varieties of programs may require introducing mutations involving substitutions of types other than integer.


%% file: results_javaUtil_short.tex
\texttt{ArrayDeque}  &  \texttt{contains}  &  $57\,${\scriptsize \%}  &  $97.95\,${\scriptsize \%}  &  7  &  14  &  (2)  &  0  &  10  &  438  &  $98\,${\scriptsize \%}  &  $39\,${\scriptsize \%}  &  $2158\,${\scriptsize s}   \\ 

\texttt{ArrayDeque}  &  \texttt{removeFirstOccurrence}  &  $53\,${\scriptsize \%}  &  $97.96\,${\scriptsize \%}  &  7  &  14  &  (2)  &  0  &  11  &  446  &  $98\,${\scriptsize \%}  &  $42\,${\scriptsize \%}  &  $2180\,${\scriptsize s}    \\ 

\texttt{ArrayDeque}  &  \texttt{removeLastOccurrence}  &  $87\,${\scriptsize \%}  &  $99.52\,${\scriptsize \%}  &  9  &  43  &  (11)  &  0  &  6  &  391  &  $93\,${\scriptsize \%}  &  $60\,${\scriptsize \%}  &  $3281\,${\scriptsize s}   \\ 

\texttt{ArrayList}  &  \texttt{clear}  &  $70\,${\scriptsize \%}  &  $95.22\,${\scriptsize \%}  &  6  &  9  &  (2)  &  0  &  6  &  26  &  $79\,${\scriptsize \%}  &  $0\,${\scriptsize \%}  &  $1524\,${\scriptsize s}   \\ 

\texttt{ArrayList}  &  \texttt{indexOf}  &  $23\,${\scriptsize \%}  &  $90.83\,${\scriptsize \%}  &  7  &  16  &  (3)  &  0  &  13  &  40  &  $89\,${\scriptsize \%}  &  $11\,${\scriptsize \%}  &  $1914\,${\scriptsize s}    \\ 

\texttt{ArrayList}  &  \texttt{lastIndexOf}  &  $20\,${\scriptsize \%}  &  $93.33\,${\scriptsize \%}  &  6  &  14  &  (2)  &  0  &  13  &  77  &  $96\,${\scriptsize \%}  &  $0\,${\scriptsize \%}  &  $1574\,${\scriptsize s}   \\ 

\texttt{ArrayList}  &  \texttt{remove}  &  $23\,${\scriptsize \%}  &  $92.87\,${\scriptsize \%}  &  7  &  16  &  (3)  &  0  &  13  &  40  &  $89\,${\scriptsize \%}  &  $13\,${\scriptsize \%}  &  $2065\,${\scriptsize s}   \\ 

\texttt{Arrays}  &  \texttt{binarySearch0}  &  $100\,${\scriptsize \%}  &  $100\,${\scriptsize \%}  &  11  &  30  &  (17)  &  0  &  6  &  1289  &  $90\,${\scriptsize \%}  &  $77\,${\scriptsize \%}  &  $4200\,${\scriptsize s}    \\ 

\texttt{Arrays}  &  \texttt{equals}  &  $100\,${\scriptsize \%}  &  $100\,${\scriptsize \%}  &  7  &  7  &  (1)  &  0  &  3  &  96  &  $80\,${\scriptsize \%}  &  $21\,${\scriptsize \%}  &  $2240\,${\scriptsize s}    \\ 

\texttt{Arrays}  &  \texttt{fill\_a}  &  $100\,${\scriptsize \%}  &  $100\,${\scriptsize \%}  &  6  &  5  &  (1)  &  0  &  1  &  6  &  $83\,${\scriptsize \%}  &  $0\,${\scriptsize \%}  &  $1391\,${\scriptsize s}   \\ 

\texttt{Arrays}  &  \texttt{fill\_b}  &  $100\,${\scriptsize \%}  &  $100\,${\scriptsize \%}  &  7  &  15  &  (5)  &  0  &  1  &  55  &  $88\,${\scriptsize \%}  &  $21\,${\scriptsize \%}  &  $1880\,${\scriptsize s}    \\ 

\texttt{Arrays}  &  \texttt{fill\_c}  &  $100\,${\scriptsize \%}  &  $100\,${\scriptsize \%}  &  6  &  7  &  (3)  &  0  &  1  &  15  &  $37\,${\scriptsize \%}  &  $53\,${\scriptsize \%}  &  $1375\,${\scriptsize s}   \\ 

\texttt{Arrays}  &  \texttt{fill\_d}  &  $100\,${\scriptsize \%}  &  $100\,${\scriptsize \%}  &  7  &  18  &  (8)  &  0  &  1  &  55  &  $70\,${\scriptsize \%}  &  $39\,${\scriptsize \%}  &  $1857\,${\scriptsize s}   \\ 

\texttt{Arrays}  &  \texttt{hashCode\_a}  &  $100\,${\scriptsize \%}  &  $100\,${\scriptsize \%}  &  2  &  4  &  (0)  &  0  &  0  &  0  &  --$\ $  &  --$\ $  &  $389\,${\scriptsize s}     \\ 

\texttt{Arrays}  &  \texttt{hashCode\_b}  &  $100\,${\scriptsize \%}  &  $100\,${\scriptsize \%}  &  2  &  4  &  (0)  &  0  &  0  &  0  &  --$\ $  &  --$\ $  &  $343\,${\scriptsize s}   \\ 

\texttt{Arrays}  &  \texttt{insertionSort\_a}  &  $100\,${\scriptsize \%}  &  $100\,${\scriptsize \%}  &  10  &  74  &  (38)  &  0  &  8  &  1224  &  $89\,${\scriptsize \%}  &  $70\,${\scriptsize \%}  &  $4029\,${\scriptsize s}   \\ 

\texttt{Arrays}  &  \texttt{insertionSort\_b}  &  $100\,${\scriptsize \%}  &  $100\,${\scriptsize \%}  &  11  &  73  &  (34)  &  0  &  8  &  5200  &  $99\,${\scriptsize \%}  &  $45\,${\scriptsize \%}  &  $4512\,${\scriptsize s}  \\ 

\texttt{Arrays}  &  \texttt{merge}  &  $0\,${\scriptsize \%}  &  $90.48\,${\scriptsize \%}  &  11  &  78  &  (62)  &  1  &  16  &  7532  &  $97\,${\scriptsize \%}  &  $8\,${\scriptsize \%}  &  $8034\,${\scriptsize s}    \\ 

\texttt{Arrays}  &  \texttt{quicksortPartition}  &  $0\,${\scriptsize \%}  &  $93.94\,${\scriptsize \%}  &  9  &  57  &  (18)  &  1  &  16  &  42714  &  $99\,${\scriptsize \%}  &  $60\,${\scriptsize \%}  &  $5657\,${\scriptsize s}    \\ 

\texttt{Arrays}  &  \texttt{vecswap}  &  $100\,${\scriptsize \%}  &  $100\,${\scriptsize \%}  &  8  &  18  &  (9)  &  0  &  5  &  1983  &  $96\,${\scriptsize \%}  &  $22\,${\scriptsize \%}  &  $2698\,${\scriptsize s}    \\ 

\texttt{Collections}  &  \texttt{replaceAll}  &  $77\,${\scriptsize \%}  &  $96.71\,${\scriptsize \%}  &  6  &  16  &  (4)  &  0  &  8  &  57  &  $77\,${\scriptsize \%}  &  $56\,${\scriptsize \%}  &  $1801\,${\scriptsize s}   \\ 

\texttt{Collections}  &  \texttt{reverse}  &  $21\,${\scriptsize \%}  &  $79.30\,${\scriptsize \%}  &  9  &  34  &  (18)  &  0  &  15  &  1181  &  $63\,${\scriptsize \%}  &  $56\,${\scriptsize \%}  &  $6949\,${\scriptsize s}    \\ 

\texttt{Collections}  &  \texttt{sort}  &  $0\,${\scriptsize \%}  &  $72.80\,${\scriptsize \%}  &  9  &  17  &  (0)  &  2  &  16  &  4343  &  $96\,${\scriptsize \%}  &  $86\,${\scriptsize \%}  &  $3933\,${\scriptsize s}    \\ 

\texttt{Vector}  &  \texttt{indexOf}  &  $100\,${\scriptsize \%}  &  $100\,${\scriptsize \%}  &  6  &  24  &  (4)  &  0  &  2  &  20  &  $70\,${\scriptsize \%}  &  $32\,${\scriptsize \%}  &  $1698\,${\scriptsize s}    \\ 

\texttt{Vector}  &  \texttt{lastIndexOf}  &  $90\,${\scriptsize \%}  &  $99.23\,${\scriptsize \%}  &  7  &  19  &  (2)  &  0  &  3  &  40  &  $88\,${\scriptsize \%}  &  $38\,${\scriptsize \%}  &  $1859\,${\scriptsize s}    \\ 

\texttt{Vector}  &  \texttt{removeAllElements}  &  $100\,${\scriptsize \%}  &  $100\,${\scriptsize \%}  &  5  &  12  &  (5)  &  0  &  1  &  10  &  $50\,${\scriptsize \%}  &  $0\,${\scriptsize \%}  &  $1218\,${\scriptsize s}   \\ 

\texttt{Vector}  &  \texttt{removeRange}  &  $63\,${\scriptsize \%}  &  $95.80\,${\scriptsize \%}  &  7  &  17  &  (5)  &  0  &  7  &  135  &  $65\,${\scriptsize \%}  &  $21\,${\scriptsize \%}  &  $2574\,${\scriptsize s}    \\ 

\texttt{Vector}  &  \texttt{setSize}  &  $100\,${\scriptsize \%}  &  $100\,${\scriptsize \%}  &  7  &  31  &  (20)  &  0  &  2  &  63  &  $58\,${\scriptsize \%}  &  $15\,${\scriptsize \%}  &  $2003\,${\scriptsize s} \\ 

\midrule 

  &  \textsc{average}  &  $71\,${\scriptsize \%}  &  $97\,${\scriptsize \%}  &  7  &  25  &  (10)  &  0  &  7  &  2410  &  $82\,${\scriptsize \%}  &  $34\,${\scriptsize \%}  &  $2691\,${\scriptsize s}    \\